\documentclass[journal]{IEEEtran}

\usepackage{amsmath}
\usepackage{amssymb}
\usepackage{array}
\usepackage{cite}
\usepackage{url}
\usepackage{arydshln}
\usepackage{graphicx}
\usepackage{mathptmx}
\usepackage{color}
\usepackage{booktabs}
\usepackage{textcomp}
\usepackage{microtype}
\usepackage{ccicons} 
\usepackage{comment}
\usepackage{xcolor}
\usepackage{nicematrix}
\usepackage[binary-units=true]{siunitx}
\usepackage{breqn}

\definecolor{flirt}{RGB}{173,0,102}
\definecolor{cerise}{RGB}{209,61,148}
\definecolor{ecstasy}{RGB}{245,133,31}
\definecolor{pear}{RGB}{202,218,42}
\definecolor{nepal}{RGB}{129,164,185}
\definecolor{mintblue}{RGB}{207,232,246}

\graphicspath{{.}{figures/}{00/}}

\begin{document}

\title{An Efficient NVoD Scheme Using Implicit Error Correction and Subchannels for Wireless Networks}

\author{Rafael Asorey-Cacheda, Antonio-Javier Garcia-Sanchez, Joan Garcia-Haro,~\IEEEmembership{Member,~IEEE}%
\thanks{Rafael Asorey-Cacheda, Antonio-Javier Garcia-Sanchez and Joan Garcia-Haro are with the Department of Information and Communication Technologies, Technical University of Cartagena, ETSIT, Campus Muralla del Mar 1, 30202 Cartagena, Spain; e-mail: \{rafael.asorey, antoniojavier.garcia, joang.haro\}@upct.es.}%
\thanks{Manuscript received date, 2019; revised date, 2019.}}

\markboth{IEEE Transactions on Multimedia,~Vol.~XX, No.~X, Month~2019}%
{Asorey-Cacheda \MakeLowercase{\textit{et al.}}: An Efficient NVoD Scheme Using Implicit Error Correction and Subchannels for Wireless Networks}

\maketitle

\begin{abstract}
Implicit Error Correction (IEC) is a near Video-on-Demand (nVoD) scheme that trades bandwidth utilization for initial playback delay to potentially support an infinite number of users. Additionally, it provides error protection without any further bandwidth increase by exploiting the implicit redundancy of nVoD protocols, using linear combinations of the segments transmitted in a given time slot. However, IEC packet loss protection is weaker at the beginning of the playback due to the lack of implicit redundancy and lower decoding efficiency, resulting in worse subjective playback quality. In tackling this issue, this paper contributes with an extension of the original nVoD architecture, enhancing its performance by adding a new element namely, subchannels. These subdivisions of the original channels do not provide further packet loss protection but significantly improve the decoding efficiency, which in turn increases playback quality, especially at the beginning. Even for very high packet loss probabilities, subchannels are designed to obtain higher decoding efficiency which results in greater packet loss protection than that provided by IEC. The proposed scheme is especially useful in wireless cooperative networks using techniques such as network coding, as content transmissions can be split into different subchannels in order to maximize network efficiency.

\end{abstract}


\IEEEpeerreviewmaketitle

\section{Introduction}

The increase of broadband Internet access has contributed to the popularization of video streaming services \cite{6587820}, and its major implications have been addressed in some works such as \cite{7128393} and \cite{7155559}. In connection to this situation, the study and development of robust techniques for reliable content distribution have become necessary. This paper presents an extended architecture of the \emph{Implicit Error Correction} (IEC) nVoD scheme that considerably improves the original proposal in \cite{mtap1} and \cite{mtap2} by significantly increasing client decoding efficiency. Specifically, we have included \emph{subchannels} to the system architecture. A subchannel is an IEC channel partition, as we will describe below. Moreover, subchannels enable IEC clients to recover from more packet losses while shortening the time required to obtain good playback quality.

There are several families of nVoD broadcast-oriented protocols such as periodic broadcast \cite{pb2X,harmonic,7782746,Altaf2017}, patching \cite{290771X,7552967,8221104}, and bandwidth skimming \cite{eager99bandwidthX,10.1007/978-981-10-7605-3_124,Abozeid:2017:SVS:3093241.3093287}. All of them split media files into segments and simultaneously broadcast them at different rates according to their relative position in the media file and the initial content playback delay. As a consequence, clients concurrently receive multiple streams with a total transmission rate proportional to the content rate. The main property of nVoD protocols is that they virtually support an unlimited number of clients with limited bandwidth at the cost of introducing an initial playback delay. However, all these protocols do not deal with packet losses, which are very common in wireless networks. In general,  packet recovery and error correction are performed transmitting more redundant data, i.e. using protocol overheads; for instance, Forward Error Correction techniques. 

Currently, most multimedia services are provisioned over broadband wired networks. However, different aspects, such as user mobility, exponential growth of wireless networks, or deployment cost of wired networks in rural areas are changing this paradigm. In Europe, solutions such as 4G/5G networks will play an important role in multimedia wireless services and should overcome problems in places where wired technology is costly \cite{940034X}. As wireless networks will grow considerably in the short term, it is important to consider diverse issues that can compromise service quality. Among them: dependence on weather or seasonal phenomena \cite{sp}; variable node density, which can significantly degrade available bandwidth; and user mobility. All of these situations can result in error rates of over 10\% \cite{er10}. Connection oriented services, such as web surfing and others, can tolerate these error rates and still provide high quality. However, in broadband multimedia services, they pose a clear challenge that has motivated research into robust coding schemes. 

In \cite{raso08} we proposed a new nVoD scheme, IEC, with all the advantages of traditional server-initiated nVoD systems, such as unlimited scalability. IEC exploits the redundancy of the systems to provide intrinsic error correction capabilities at the client end without extra bandwidth cost. In \cite{mtap1} we evaluated this scheme, both analytically and by means of computer simulation, demonstrating its feasibility. Additionally, in \cite{5425298} we studied IEC in wireless cooperative networks using network coding. This work showed that IEC obtains significant benefits in terms of protection against packet losses and bandwidth utilization of the joint coding schema.   
 
Related to this research, there have also been studies dealing with nVoD and error protection. The work described in \cite{jenkac} and \cite{Jenkac2006} is based on harmonic nVoD schemes. This approach splits content into segments of equal size, internally divides segments into chunks, and applies linear coding at different rates. To achieve error protection, it produces longer coded segments; i.e. adds redundancy that can be used in case of packet losses. As a result, protection is equal for all segments at the cost of adding an initial playback delay that rises exponentially with protection level. At this point, it is important to stress that protection levels for the schemes in \cite{jenkac} and \cite{Jenkac2006} remain constant during the entire playback, while the IEC scheme, due to its architecture, increases it every time a new slot is downloaded; i.e. over time.

Regarding content distribution and wireless networks, recent works deal with the paradigm of the Internet of Things and optimal content distribution, such as in \cite{10.9717/JMIS.2018.5.1.27,MEMOS2018619,STERGIOU2018964,PLAGERAS2018349}. If we focus on VoD and wireless networks, some of which are related to content distribution in mobile networks, such as \cite{Psannis2006}, in which authors proposed the utilization of different multicast encoded versions of the streams to distribute video in 3G wireless networks.

Some works also propose the utilization of network coding for VoD streaming. In this regard, there are several investigations in the academic literature that deal with these two topics, such as \cite{6416071,6600846,8013842}. The application of network coding is beyond the scope of this paper, but IEC can be easily adapted to this transmission paradigm, as described in \cite{5425298}. Basically, network coding is based on forwarding packets towards their destination/s through each network node. Thus, these intermediate nodes process packets in addition to relaying them. Processing packets consists of generating and sending new packets as linear combinations of the incoming ones. As a result, network coding can increase error resilience in a similar way to the systems described in the paragraph above.

Other papers related to VoD and wireless networks recommend the utilization of scalable video coding (SVC) to adapt the stream bandwidth (i.e. stream quality) to the number of users \cite{6410040}. SVC has many advantages, such as better network scalability, better adaptability to different devices, or custom protection of different video layers. Moreover, SVC  can also be implemented in nVoD systems, such as IEC, which was described in \cite{5418877}, and combined with other complementary techniques, such as channel or network coding.

This paper is based on previous works \cite{mtap1} and \cite{mtap2}. The work in \cite{mtap1} introduced the basic IEC scheme and analyzed its performance in terms of packet loss recovery. The work in \cite{mtap2} showed an implementation architecture of IEC and extended the original IEC mechanism by introducing a new technique, denoted \textit{Feedback Error Correction}, which increases the packet loss recovery capabilities.

The main contributions of this paper are summarized in the following points:
\begin{itemize}
	\item An extension of the original IEC nVoD architecture to support subchannels as an effective method of improving decoding efficiency. The advantages of using subchannels are at the cost of increased encoding and decoding complexity. However, as the original IEC nVoD computational complexity was very small \cite{mtap2} and the number of subchannels required to achieve good performance is also low, this complexity is constrained and should not be considered as a limiting factor.
	\item A significant improvement achieved in decoding efficiency compared to that of previous IEC variants, which is more noticeable at the beginning of the video playback.  
	\item A mid and long-term enhancement in IEC packet loss protection as a consequence of the improvement of decoding efficiency.
	\item An improvement in video playback quality at the client end.
\end{itemize}

The paper is organized as follows. Firstly, Section \ref{iec} describes the IEC nVoD scheme and the improvements provided by redundancy and Feedback Error Correction, which is part of the advanced IEC implementation.  Section \ref{sub} presents and analyzes the inclusion of subchannels in the IEC architecture. In Section \ref{num}, we validate our results with some numerical tests. Finally, we conclude the paper and propose future lines in Section \ref{conclusions}.

\section{The IEC nVoD scheme}\label{iec}

The basic IEC schema was introduced in \cite{raso08}, and is based on \textit{Harmonic Broadcasting} (HB) algorithms \cite{juhn97harmonicX,juhn2X,8012499}. The implementation architecture of IEC was presented in \cite{mtap1} and an advanced IEC schema, incorporating Feedback Error Correction, was described in \cite{mtap2}. For the sake of clarity, this section summarizes the main properties and features of the IEC schema. More detailed explanations can be found in \cite{raso08}, \cite{mtap1}, and \cite{mtap2}.

\begin{figure}[tp]
\centering
\includegraphics[scale=0.75]{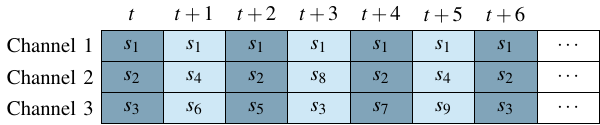}
\caption{Illustrative example of a suboptimal HB scheme (constant number of channels per time slot and a single segment size).}
\label{fig1}
\end{figure}

HB algorithms split contents into segments ($s_1,s_2,\ldots,s_n$, $n \equiv$ number of segments) of equal size. Each segment is composed of $k$ packets and transmitted using $M \in \mathbb{N}$ independent multicast channels at different transmission frequencies or periods, so that transmissions are organized in time slots with a duration equal to the duration of a segment\footnote{A segment frequency or, equivalently, a segment period sets a deadline for segment retransmission. For example, a segment with period $1$ must be transmitted in all time slots and a segment with period $2$ must be transmitted every two time slots.}. Moreover, in order to minimize bandwidth utilization, segment periods must be as long as possible. Figure \ref{fig1} shows an example of an HB (\textit{suboptimal}) scheme. This HB scheme will be utilized as a reference for all subsequent examples appearing in this paper. In Figure \ref{fig1}, rows correspond to channels and columns to time slots. Thus, as can be observed, segment $s_1$ has a retransmission frequency of $1$, segment $s_2$ has a retransmission frequency of at least $1/2$, segment $s_3$ has a retransmission frequency of at least $1/3$ and, in general, an arbitrary segment $s_n$ has a retransmission frequency of at least $1/n$. As the number of channels is fixed, bandwidth utilization is constant but suboptimal. As a consequence, any \textit{regular scheme} must satisfy $\mathrm{T}(s_i) \leq i$. ($\mathrm{T}(s_i)$ is the period of segment $s_i$) This constraint establishes the minimum required bandwidth, but any regular nVoD scheme might additionally allocate bandwidth for redundancy if it is available. Thus, let us define $M$ as the minimum number of channels required for the nVoD scheme and $R$ as the number of redundancy channels. Consequently, the total number of channels of the nVoD scheme, $I$, is obtained as $I=M+R$.

The minimum bandwidth, $B_n$, required by the IEC scheme can be derived from the constraint $\mathrm{T}(s_i) \leq i$. Thus:

\begin{equation}
B_n = \sum_{i=1}^n \frac{1}{\mathrm{T}(s_i)} r,
\end{equation}

\noindent where $r$ is the content rate. Consequently, the minimum $B_n$ corresponds to $T(s_i) = i$. Therefore,

\begin{equation}
B_n \geq \sum_{k=1}^n \frac{1}{k} r = H_n r,
\end{equation}

\noindent where $H_n$ is a harmonic number defined by:

\begin{equation}\label{hn}
H_n = \sum_{k=1}^n \frac{1}{k}
\end{equation}

Expression \eqref{hn} implies that $M \geq H_n$, $n \geq 1$ (harmonic numbers are always non-integers, except for $n=1$). However, $M \gtrapprox H_n$ as $n$ grows. For this reason, it is advisable to choose an $n$ so that $M \approx H_n$, $M \in \mathbb{N}$ (the difference is bounded by $1/n$).

\begin{figure}[tp]
\centering
\includegraphics[scale=0.75]{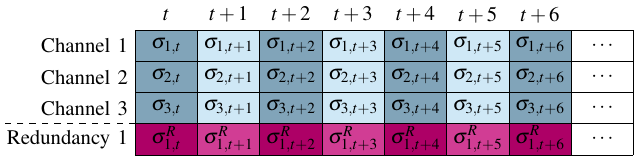}
\caption{The same scheme as in Figure \ref{fig1} with implicit error correction and a redundancy channel.}
\label{fig3}
\end{figure}

As explained in \cite{raso08}, content is divided into $n$ equal size segments, $\mathcal{S} = \{s_1, s_2, \ldots, s_n\}$. The encoded segments $\sigma_{m,t}$ for channel $m$ and time slot $t$ (Figure \ref{fig3}) are obtained by performing the following operation on the subset of original segments $s_i \ldots s_j$ ($|\{s_i \ldots s_j\}| = M \geq H_n$):

\begin{equation}\label{exp1}
\sigma_{m,t} = c_{s_{i}}^{m,t}\cdot s_{i} \oplus \ldots \oplus c_{s_{j}}^{m,t}\cdot s_{j},\,
\{c_{s_{i}}^{m,t} \ldots c_{s_{j}}^{m,t}\} \in \{0,1\},
\end{equation}

\noindent where ``$\oplus$'' is the standard bitwise XOR operation, ``$\cdot$'' is the multiplication, and $\{c_{s_{i}}^{m,t} \ldots c_{s_{j}}^{m,t}\}$ are coefficients, suitably chosen to maximize decoding efficiency \cite{mtap2}. These coefficients indicate whether a segment $s_i$ is going to be used for a bitwise XOR operation ($c_{s_{i}}^{m,t} = 1$). For example, in slot $t$ of the example shown in Figure \ref{fig1}, the transmitted segments are $s_1$, $s_2$, and $s_3$ and one possible $\sigma_{1,t}$ could be $\sigma_{1,t}=s_1 \oplus s_3$.

As indicated above, IEC supports the addition of an arbitrary number $R$ of redundancy channels, $R \geq 0$. As in expression \eqref{exp1}, a redundancy channel segment can be expressed as:

\begin{equation}\label{expx}
\sigma_{m,t}^R = \rho_{s_{i}}^{m,t}\cdot s_{i} \oplus \ldots \oplus \rho_{s_{j}}^{m,t}\cdot s_{j},\, 
\{\rho_{s_{i}}^{m,t} \ldots \rho_{s_{j}}^{m,t}\} \in \{0,1\}
\end{equation}

In expression \eqref{expx}, coefficients $\{\rho_{s_{i}}^{m,t} \ldots \rho_{s_{j}}^{m,t}\}$ are equivalent to those of expression \eqref{exp1}. Let $C^t$ be a binary matrix of size $I \times M$ whose components are:

\begin{equation}\label{ct}
C^t=
\left[\begin{NiceArray}{ccc}
c^{1,t}_{s_i} & \Cdots & c^{1,t}_{s_j} \\
c^{2,t}_{s_i} & \Cdots & c^{2,t}_{s_j} \\
\Vdots & \Ddots & \Vdots \\
c^{M,t}_{s_i} & \Cdots & c^{M,t}_{s_j} \\ \hdashline[1pt/1pt]
\rho^{1,t}_{s_i} & \Cdots & \rho^{1,t}_{s_j}\\
\Vdots & \Ddots & \Vdots \\
\rho^{R,t}_{s_i} & \Cdots & \rho^{R,t}_{s_j}
\end{NiceArray}\right]_{I \times M}
\end{equation}

As can be expected, for any time slot $t$, the rank of $C^t$ must be $M$ to recover all the segments. This issue is not studied in this paper but has been discussed in previous works \cite{raso08,mtap1,mtap2}.

It is important to note that a segment is composed of an arbitrary number of packets. Depending on the error distribution, it will be possible to recover parts of a segment even if others are lost or recovery is not feasible.  

\subsection{IEC peformance}

The work in \cite{mtap2} provided extensive information on the architecture of the IEC decoding system. In short, if some coded segments are lost, the recovery process can use previously decoded segments and the redundancy channels to recover these lost segments. This permits the correction of errors up to a certain packet loss probability. However, this probability always increases if the client decodes a new segment (or a packet).

Regarding packet loss probability, the works in \cite{raso08}, \cite{mtap1}, and \cite{mtap2} only considered uniform distributions. It can be argued that other packet loss distributions are possible, such as bursty packet losses, especially in wireless communications, and consequently providing different performance metrics. However, as packets within a time slot do not need to be transmitted in order, their transmission sequence can be randomized and any type of packet loss distribution will be transformed into a uniform one at the receiving end \cite{592604}.

In \cite{raso08}, the formulation of the maximum admissible packet loss rate as downloading progresses was first introduced. Let us define $p_{n,b,R}$ as the maximum admissible packet loss probability for an IEC scheme with $n$ segments, $H_n \approx M$, during time slot $b$ using $R$ channels of redundancy, $I=M+R$. Thus, to decode all the segments in a time slot $b$, equation \eqref{pnbr} must hold:

\begin{equation}\label{pnbr}
p_{n,b,R} \leq \frac{H_{b-1}+R}{H_n + R}
\end{equation}

\begin{figure}[tp]
\centering
\includegraphics{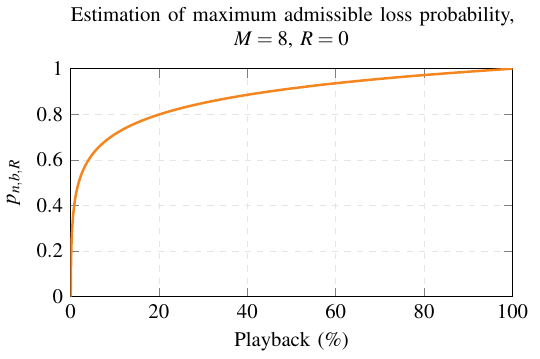}
\caption{Maximum admissible $p_{n,b,R}$ as new original segments become available for $M=8$ and $R=0$.}
\label{pnb}
\end{figure}

To illustrate expression \eqref{pnbr}, Figure \ref{pnb} represents an example of the maximum admissible packet loss probability as new segments are decoded and become available. This example employs no redundancy, $R=0$, and $M=8$ multicast channels. As can be seen, even without redundancy channels probability $p_{n,b,0}$ increases as new segments, obtained in previous time slots, become available.

\begin{figure}[tp]
\centering
\includegraphics{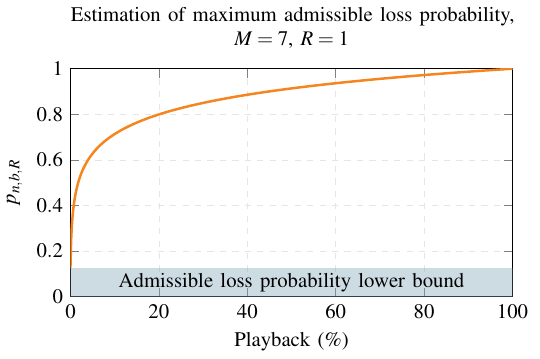}
\caption{Maximum admissible $p_{n,b,R}$ as new original segments become available for $M=7$ and $R=1$.}
\label{pnb-R}
\end{figure}

Expression \eqref{pnbr} has a global minimum at $b=1$ because $H_n$ is a monotonically increasing unbounded function. Thus, from expression \eqref{pnbr} it is evident that $p_{n,1,R} = R/(H_n+R)$ is the minimum admissible packet loss probability for any time slot. As stated above, IEC implicit channel redundancy, i.e. previously decoded segments, increases the admissible packet loss rate in future time slots. As an example, Figure \ref{pnb-R} shows $p_{n,b,R}$ for an IEC scheme with one channel of redundancy, $R=1$, and $M=7$ multicast channels. This channel of redundancy provides minimum packet loss protection of up to 12.5\%. However, as a drawback, this approach increases the initial playback delay. This happens because using redundancy channels decreases $M$ (there are fewer channels to assign segments) and consequently, decreases $H_n$ (note that $H_n \leq M$ must hold).

It is possible to estimate a relationship between the number of channels and delay. In fact, harmonic numbers can be approximated by $H_n = \gamma + \ln n$ if $n$ is large enough, where $\gamma$ is the constant of Euler-Mascheroni ($\gamma \approx 0.577$). Thus, as explained in \cite{mtap2}, increasing the number of redundancy channels (mantaining $I = M+R$ constant) decreases the number of segments that can be allocated in the remaining channels (i.e. segment size increases) by a factor of $e^R$. As a consequence, the utilization of $R$ redundancy channels increases delay proportionally to $e^R$.

\begin{figure}[tp]
\centering
\includegraphics{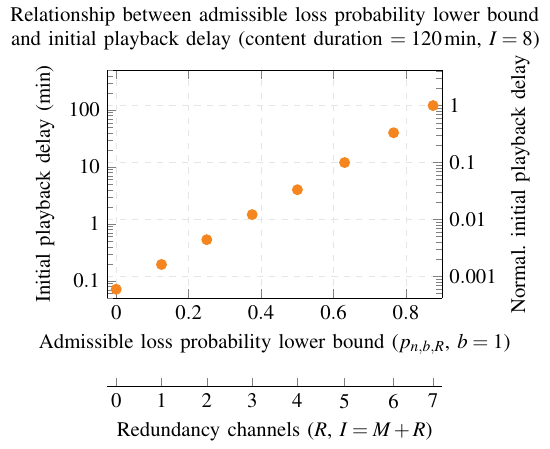}
\caption{Relationship between admissible lower bound ($p_{n,b,R}$, $b=1$) and initial playback delay for content duration=$\SI{120}{\minute}$ and $I=8$.}
\label{fignew:1}
\end{figure}

Figure \ref{fignew:1} is an illustrative example of the relationship between admissible loss probability lower bound and initial playback delay for content with a duration of \SI{120}{\minute} and $I=8$ channels (content and redundancy channels). As predicted by our theory, initial playback delay grows exponentially with the utilization of transmission channels as redundancy channels. However, it should be noted that for a content duration of \SI{120}{\minute}, an admissible loss probability lower bound of 25\% ($R=2$) imposes fewer than \SI{32}{\second} of initial playback delay. The lowest initial playback delay is achieved for an admissible loss probability lower bound of 0 with an initial playback delay of fewer than \SI{5}{\second}.

Thus, from expression \eqref{pnbr} the following can be derived:

\begin{equation}
\label{13}
p_{n',b',R} \approx \frac{H_{b-1}}{H_n} = p_{n,b,0},\, n \approx n'e^R,\, (b-1) \approx (b'-1)e^R
\end{equation}

\begin{figure}[tp]
\centering
\includegraphics{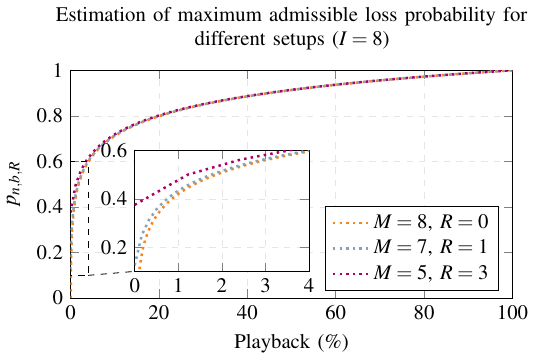}
\caption{Maximum admissible $p_{n,b,R}$ as new original segments become available for different setups ($I=8$).}
\label{fignew:2}
\end{figure}

Expression \eqref{13} reflects that redundancy channels do not improve packet loss recovery during playback. Nevertheless, redundancy provides minimum packet loss protection at the cost of increasing initial playback delay as can be observed in Figure \ref{fignew:1}. Figures \ref{pnb-R} and \ref{fignew:1} show how initial playback delay is increased approximately by a factor of $e$ compared to Figure \ref{pnb}. Figure \ref{fignew:2} compares the maximum admissible loss probability for different setups. In accordance with expression \eqref{13}, these values are very similar and converge as playback grows. However, redundancy channels raise the minimum admissible packet loss probability as depicted in Figure \ref{fignew:1}. If IEC is compared to the HB nVoD scheme in \cite{jenkac} and \cite{Jenkac2006}, using the same setup of initial playback delay and bandwidth utilization, the theoretical initial packet loss protection is the same\footnote{Assuming ideal conditions, such as perfect decoding efficiency and uniform packet loss distribution.}. However, packet loss protection in \cite{jenkac} and \cite{Jenkac2006} remains the same during playback because it depends on the redundancy added to each segment, whereas IEC utilizes redundancy channels in a similar way as that in \cite{jenkac} and \cite{Jenkac2006} and, in addition, previously downloaded segments act as extra redundancy and gradually increase packet loss protection. In other words, IEC provides at least the same packet loss protection as \cite{jenkac} and \cite{Jenkac2006}, plus the protection provided by the previously downloaded segments.

Moreover, it can be argued that nVoD schemes are inefficient in terms of bandwidth allocation if the number of clients is low. Although this may be true, if an IEC server is provided with information about clients, it can selectively switch off some channels, as clients have implicit redundancy from previously downloaded segments for successful decoding.

\subsection{IEC architecture}

The work in \cite{mtap2} introduced an advanced architecture implementation of IEC and further included Feedback Error Correction (FEC).

\begin{figure}[tp]
\centering
\includegraphics[scale=0.7]{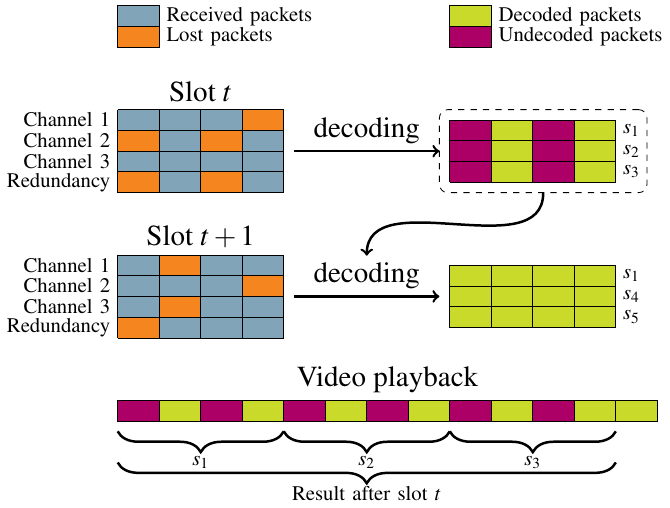}
\caption{Decoding example without FEC.}
\label{fec1}
\end{figure}

The IEC implementation described in \cite{mtap1} worked as follows (see Figure \ref{fec1}):

\begin{enumerate}
	\item During a time slot, the decoder tries to decode received packets. If packet losses are below the admissible packet loss, all information will be recovered. Over the admissible packet loss, it will not be possible to recover all the original information.
	\item After step 1, the decoder tries to use previously downloaded packets that can act as redundancy in the decoding process.
	\item In the last state, all undecoded information is considered useless and, consequently, discarded.
\end{enumerate}

\begin{figure*}[tp]
\centering
\includegraphics[scale=0.7]{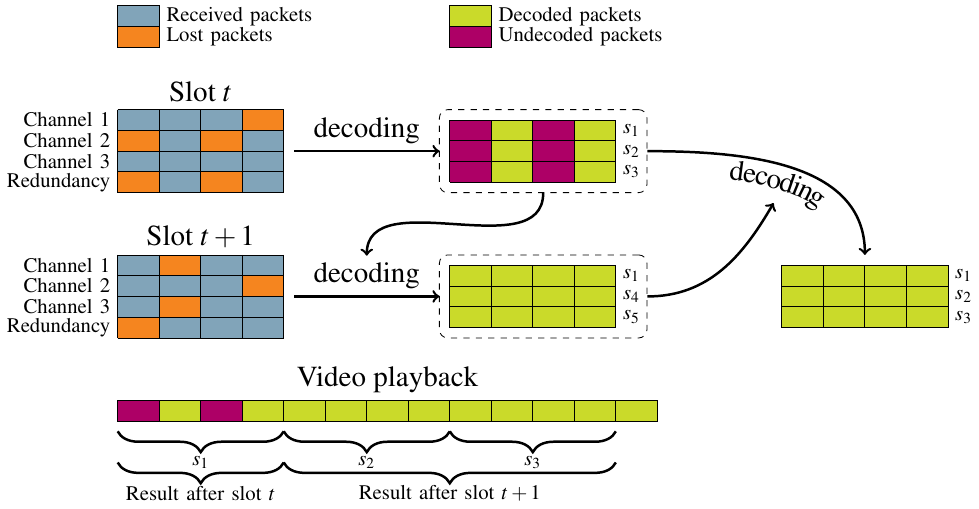}
\caption{Decoding example with FEC.}
\label{fec2}
\end{figure*}

FEC is an addition to the basic IEC scheme that takes advantage of previously undecoded information (see Figure \ref{fec2}):

\begin{enumerate}
	\item Undecoded packets are not discarded anymore but stored for future decoding. It might happen that some of these undecoded packets are decoded after their playback deadline. Even if this is the case, no information is discarded.
	\item At the end of every time slot, the decoder also tries to decode these previously undecoded packets using the newly decoded packets of the current time slot.
\end{enumerate}

The main advantage of this approach is that storing undecoded packets and trying to decode them in the subsequent time slots improves decoding performance. Once these packets belonging to previous time slots have been decoded, IEC behavior substantially improves in two ways:

\begin{enumerate}
	\item It is possible that a packet is recovered after its playback instant. However, it improves playback quality because it helps to increase the admissible packet loss in the following time slots. In other works, these types of packets can be used as redundancy in future time slots.
	\item If a packet has not been played, beyond improving the admissible packet loss in subsequent time slots, it also enhances subjective playback quality, which is the most important goal of IEC performance.
\end{enumerate}

\begin{figure}[tp]
\centering
\includegraphics{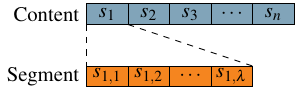}
\caption{Content segmentation prior to transmission.}
\label{exfig}
\end{figure}

\begin{figure}[tp]
\centering
\includegraphics[scale=0.75]{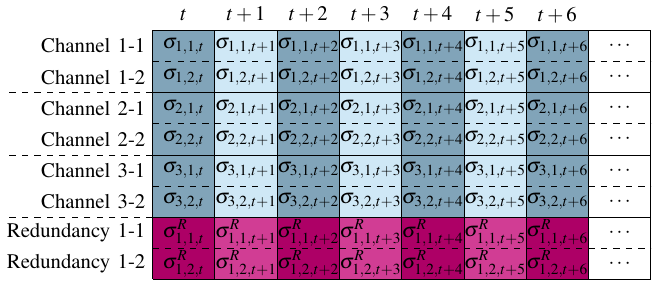}
\caption{Same scheme as in Figure \ref{fig3} but with subchannels and redundancy.}
\label{fig6}
\end{figure}

FEC, as presented in \cite{mtap2}, does not discard undecoded packets after a decoding process. These undecoded segments are stored for future decoding. This means that the IEC decoder tries to decode previously undecoded packets at every time slot, thanks to which new packets will be decoded and used as implicit redundancy. The FEC IEC implementation of \cite{mtap2} is outlined in Figure \ref{fec2}. It shows how FEC improves decoding efficiency regarding the analogous example of Figure \ref{fec1}. Going back to the same example from the previous paragraph, it can be seen that the new parts of $s_1$ decoded during $t+1$ allow the complete decoding of segments $s_2$ and $s_3$. Thus, decoding efficiency and packet loss recovery improve through the use of FEC.

\section{IEC and subchannels}\label{sub}

IEC packet loss protection is provided by redundancy channels and implicit redundancy from previously downloaded information. This is achieved by means of channel coding. As in other HB schemes, a reduced number of channels is enough to provide low initial playback delays, which is a desired property in order to optimize bandwidth utilization. However, this has a major drawback, especially in the first time slot, because packet loss protection is more efficient if the number of coded packets is higher. As an example, let us consider $I = M+R=8$, $M=7$, $R=1$, a uniform packet loss probability $p_e=0.1$, and the first time slot transmission ($b=0$). This setup provides packet loss protection for the first time slot of up to $12.5\%$, which should be enough for the system to work properly. In this scenario, the transmission involves sending an amount of sets composed of 8 coded packets, each one corresponding to a channel. The amount of sets depends on the segment size. Thus, if two or more packets of the set get lost, no decoding will be possible, as packet losses will exceed the packet loss protection provided by IEC. By means of simple probability calculations, it is easy to check that about $19\%$ of the sets will not be decoded; i.e. during playback, about $19\%$ of the first segment will contain errors. 

As described in \cite{raso08}, \cite{mtap1}, and \cite{mtap2}, in general, first content segments are the ones containing more errors. Our proposal to overcome this problem and improve playback quality, specially at the beginning, is the addition of subchannels to the IEC scheme. Subchannels do not provide better packet loss protection but better decoding efficiency, as is explained below.

The original IEC scheme uses several channels to simultaneously transmit the scheduled segments in a time slot, and each segment occupies one channel. To extend this scheme, we propose the utilization of subchannels; i.e. a segment is split into subsegments that are then transmitted by several subchannels during a time slot. The set of subchannels assigned to a segment constitutes a channel. From now on, we will refer to this scheme as IEC-S.

Let us split a segment $s_i$ into $\lambda$ equal size subsegments. Each subsegment corresponds to a subchannel with a transmission rate $r/\lambda$, such as $s_i = s_{i,1} \cup s_{i,2} \cup \ldots \cup s_{i,\lambda}$ (see example in Figure \ref{exfig}). Then, expression \eqref{exp1} becomes:

\begin{dmath}\label{exp1-s}
\sigma_{m,\mu,t} = c_{s_{i,1}}^{m,\mu,t}\cdot s_{i,1} \oplus \ldots \oplus c_{s_{i,\lambda}}^{m,\mu,t}\cdot s_{i,\lambda} 
\oplus \ldots \oplus c_{s_{j,1}}^{m,\mu,t}\cdot s_{j,1}
\oplus \ldots \oplus c_{s_{j,\lambda}}^{m,\mu,t}\cdot s_{j,\lambda},\,
\{c_{s_{i,1}}^{m,\mu,t} \ldots c_{s_{j,\lambda}}^{m,\mu,t}\} \in \{0,1\},
\end{dmath}

\noindent where $\sigma_{m,\mu,t}$ is the encoded segment for channel $m$, subchannel $\mu$ and time slot $t$. Similarly to expression \eqref{exp1}, equation \eqref{expx} becomes:

\begin{dmath}\label{expx-x}
\sigma_{m,\mu,t}^R = \rho_{s_{i,1}}^{m,\mu,t}\cdot s_{i,1} \oplus \ldots \oplus \rho_{s_{i,\lambda}}^{m,\mu,t}\cdot s_{i,\lambda} \oplus \ldots \oplus \rho_{s_{j,1}}^{m,\mu,t}\cdot s_{j,1}
\oplus \ldots \oplus \rho_{s_{j,\lambda}}^{m,\mu,t}\cdot s_{j,\lambda},\,
\{\rho_{s_{i,1}}^{m,\mu,t} \ldots \rho_{s_{j,\lambda}}^{m,\mu,t}\} \in \{0,1\},
\end{dmath}

In expressions \eqref{exp1-s} and \eqref{expx-x}, $c_{s_{i,x}}^{m,\mu,t}$ and $\rho_{s_{i,x}}^{m,\mu,t}$, $x \in \{1, \ldots, \lambda\}$ are coefficients, carefully chosen to maximize decoding efficiency \cite{mtap1}, that denote whether a subsegment $s_{i,x}$ is going to be used for a bitwise XOR operation ($c_{s_{i,x}}^{m,\mu,t} = 1$ or $\rho_{s_{i,x}}^{m,\mu,t} = 1$). Figure \ref{fig6} shows the same example as Figure \ref{fig3} but with two subchannels ($\lambda=2$) for each segment transmission. The coding matrix $C^t$ will have a size of $\lambda (M+R)$ rows by $\lambda M$ columns:

\begin{equation}\label{ctext}
C^t=\left[
\begin{NiceArray}{ccccccc}
c^{1,1,t}_{s_{i,1}} & \Cdots & c^{1,1,t}_{s_{i,\lambda}} & \Cdots &  c^{1,1,t}_{s_{j,1}} & \Cdots & c^{1,1,t}_{s_{j,\lambda}} \\
\Vdots & \Ddots & \Vdots & \Ddots & \Vdots & \Ddots & \Vdots \\
c^{1,\lambda,t}_{s_{i,1}} & \Cdots & c^{1,\lambda,t}_{s_{i,\lambda}} & \Cdots & c^{1,\lambda,t}_{s_{j,1}} & \Cdots & c^{1,\lambda,t}_{s_{j,\lambda}}\\
\Vdots & \Ddots & \Vdots & \Ddots & \Vdots & \Ddots & \Vdots \\
c^{M,\lambda,t}_{s_{i,\lambda}} & \Cdots & c^{M,\lambda,t}_{s_{i,\lambda}} & \Cdots & c^{M,\lambda,t}_{s_{j,\lambda}} & \Cdots & c^{M,\lambda,t}_{s_{j,\lambda}} \\ \hdashline[1pt/1pt]
\rho^{1,1,t}_{s_{i,1}} & \Cdots & \rho^{1,1,t}_{s_{i,1}} & \Cdots & \rho^{1,1,t}_{s_{j,1}} & \Cdots & \rho^{1,1,t}_{s_{j,1}} \\
\Vdots & \Ddots & \Vdots & \Ddots & \Vdots & \Ddots & \Vdots \\
\rho^{1,\lambda,t}_{s_{i,\lambda}} & \Cdots & \rho^{1,\lambda,t}_{s_{i,\lambda}} & \Cdots & \rho^{1,\lambda,t}_{s_{j,\lambda}} & \Cdots & \rho^{1,\lambda,t}_{s_{j,\lambda}} \\
\Vdots & \Ddots & \Vdots & \Ddots & \Vdots & \Ddots & \Vdots \\
\rho^{R,\lambda,t}_{s_{i,\lambda}} & \Cdots & \rho^{R,\lambda,t}_{s_{i,\lambda}} & \Cdots & \rho^{R,\lambda,t}_{s_{j,\lambda}} & \Cdots & \rho^{R,\lambda,t}_{s_{j,\lambda}}
\end{NiceArray}\right]_{\lambda I \times \lambda M}
\end{equation}

\begin{figure}[tp]
\centering
\includegraphics{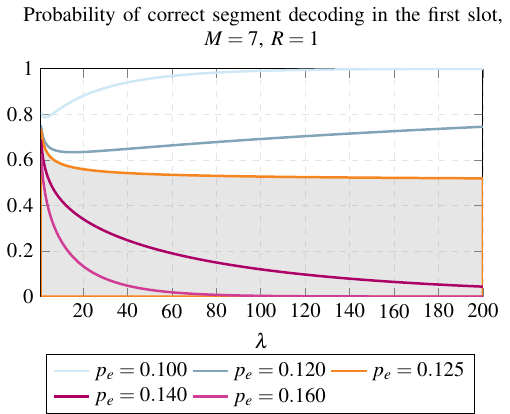}
\caption{Theoretical efficiency in the first slot as a function of $\lambda$ and $p_e$, $M=7$, $R=1$.}
\label{pnb-pr}
\end{figure}

The coding matrix of expression \eqref{ctext} is similar to that of expression \eqref{ct} but $\lambda^2$ larger. Thus, from the point of view of packet coding and decoding, the coding matrix \eqref{ctext} would be equivalent to that of an IEC scheme using a total of $\lambda I$ channels, of which $\lambda R$ are redundancy channels. Thus, every row of \eqref{ctext} represents the coding operations performed for each one of the subchannels, whereas every column of \eqref{ctext} indicates whether a subsegment is used in a subchannel coding operation.

\begin{figure}[tp]
\centering
\includegraphics{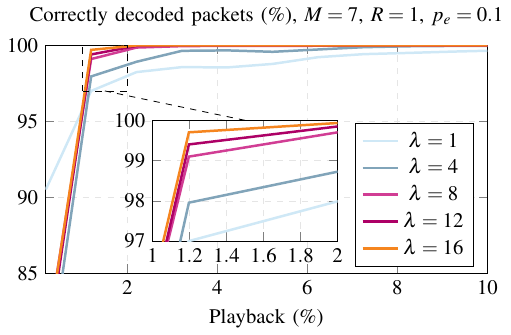}
\caption{Simulation results for different numbers of subchannels, $M=7$, $R=1$, $p_e=0.1$.}
\label{conred}
\end{figure}

\begin{figure}[tp]
\centering
\includegraphics{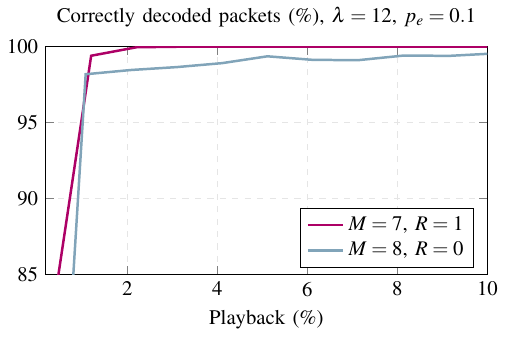}
\caption{Simulation results with and without redundancy, $I=8$, $p_e=0.1$.}
\label{sinred}
\end{figure}

\subsection{IEC-S performance}

The coding matrix \eqref{ctext} must be carefully selected to provide high packet loss probability. A bad coding matrix will fail to decode even with packet losses below the admissible packet loss probability. This issue was already discussed in previous works such as \cite{mtap2}. The work in \cite{Blomer95anxor-based} also deals with the properties required by these coding matrices. Similarly, as discussed above, let us define $p_{n,b,R,\lambda}$ as the maximum admissible packet loss probability for an IEC-S scheme with $n$ segments during time slot $b$ using $R$ channels of redundancy, where each channel is divided into $\lambda$ subchannels. In order to decode all the segments in time slot $b$, the following must hold:

\begin{equation}\label{pnbrl}
(1 - p_{n,b,R,\lambda})\lambda(H_n + R) + \lambda H_{b-1} \geq \lambda H_n \Rightarrow 
p_{n,b,R,\lambda} \leq \frac{H_{b-1}+R}{H_n + R}
\end{equation}

\begin{figure}[tp]
\centering
\includegraphics{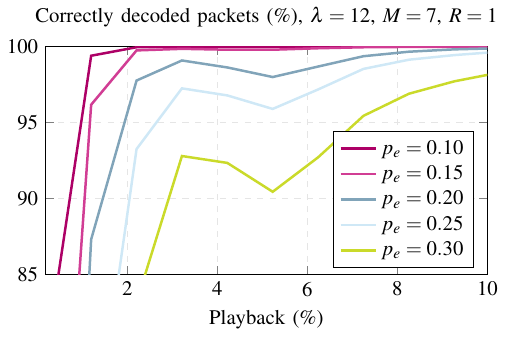}
\caption{Simulation results for different packet loss probabilities, $M=7$, $R=1$.}
\label{alto}
\end{figure}

Expression \eqref{pnbrl} demonstrates that the number of subchannel packets received ($\lambda(M+R) \approx \lambda(H_n+R)$) plus the implicit redundancy from previous time slots ($\lambda H_{b-1}$) must be equal to or higher than the number of subchannels ($\lambda M \approx \lambda H_n$). This equation shows that subchanneling does not increase the maximum admissible packet loss probability. However, let us focus on the first time slot and assume a packet loss probability of $p_e$, and that $I$ channels consisting of $\lambda$ subchannels are available.

\begin{enumerate}
\item A client correctly decodes all channels if at least $\lambda M$ channels are received. This probability can be calculated as:

\begin{equation}
\sum_{i=\lambda M}^{\lambda(M+R)} \binom{\lambda(M+R)}{i} (1-p_e)^i p_e^{\lambda(M+R)-i}
\end{equation}

\item The probability of success in the first time slot can be modeled as a binomial distribution, $\mathcal{X} \sim \mathrm{B}(\lambda (M+R), 1-p_e)$, where $f(i; \lambda(M+R), 1-p_e)$ is the probability mass function. Thus, for any $\lambda \geq 1$, we can generalize the expression for the probability of success as:

\begin{equation}\label{pr}
\mathrm{Pr}[\mathcal{X}_\lambda \geq \lambda M] = \sum_{i=\lambda M}^{\lambda(M+R)} f(i; \lambda(M+R), 1-p_e)
\end{equation}

\item By means of simple experiments it is easy to assess that subchanneling may result in lower success probabilities for the first time slot for low values of $\lambda$. However, this changes for subsequent time slots. If $\lambda$ is large enough, decoding efficiency will improve. Following the example at the beginning of this Section, for $\lambda=6$, $20\%$ of the first segment will contain errors (larger than for $\lambda=1$), and for $\lambda=12$, $16\%$ of the first segment will contain errors (lower than for $\lambda=1$). 

$\mathrm{Pr}[\mathcal{X}_\lambda \geq \lambda M]$ converges asymptotically to three different values depending on the value of $p_e$ as $\lambda \rightarrow \infty$:

\begin{equation}\label{23}
\lim \limits_{\lambda \to \infty } \mathrm{Pr}[\mathcal{X}_\lambda \geq \lambda M]  = \begin{cases}
1, &\text{if $(1-p_e)(M+R) > M$}\\
\displaystyle{\frac{1}{2}}, &\text{if $(1-p_e)(M+R) = M$}\\
0, &\text{if $(1-p_e)(M+R) < M$}
\end{cases}
\end{equation}

\end{enumerate}

\begin{figure}[tp]
\centering
\includegraphics{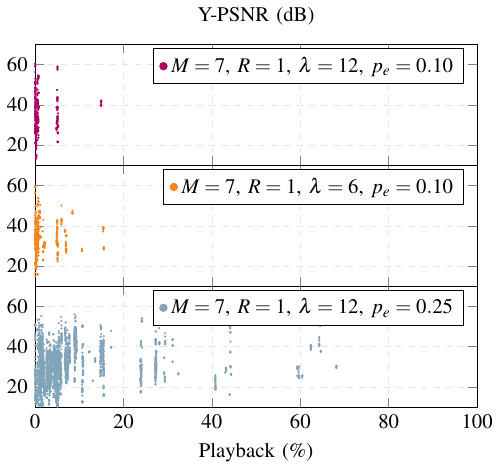}
\caption{Y-PSNR for three different setups.}
\label{psnr}
\end{figure}

Expression \eqref{23} reveals that if there is enough redundancy to support packet losses, subchannels will improve their efficiency and allow full error recovery. Of course, if $p_e$ is too high, no information will be recovered, even for high values of $\lambda$. As we will describe in Section \ref{num}, the implicit redundancy provided by IEC-S helps achieve full recovery, even for values of $p_e$ that would lead to poor results in other schemes, such as \cite{jenkac} and \cite{Jenkac2006}. 

Figure \ref{pnb-pr} represents the theoretical efficiency in the first slot for different values of $p_e$ as a function of $\lambda$ for $M=7$, $R=1$. We can observe three different behaviors depending on whether $p_e$ is greater than, equal to, or lower than 0.125. However, we will show that IEC can provide good performance even for values of $p_e$ falling within the shaded region in Figure \ref{pnb-pr} ($\lim \limits_{\lambda \to \infty } \mathrm{Pr}[\mathcal{X}_\lambda \geq \lambda M] = 0$).

\begin{figure*}[tbp!]
\centering
\setlength{\tabcolsep}{2pt}
\begin{tabular}{m{0.46\textwidth}m{0.46\textwidth}m{0.05\textwidth}}
\hspace{0.23\textwidth}\textbf{\small (a)} & \hspace{0.23\textwidth}\textbf{\small (b)} &\\
\includegraphics[width=0.46\textwidth]{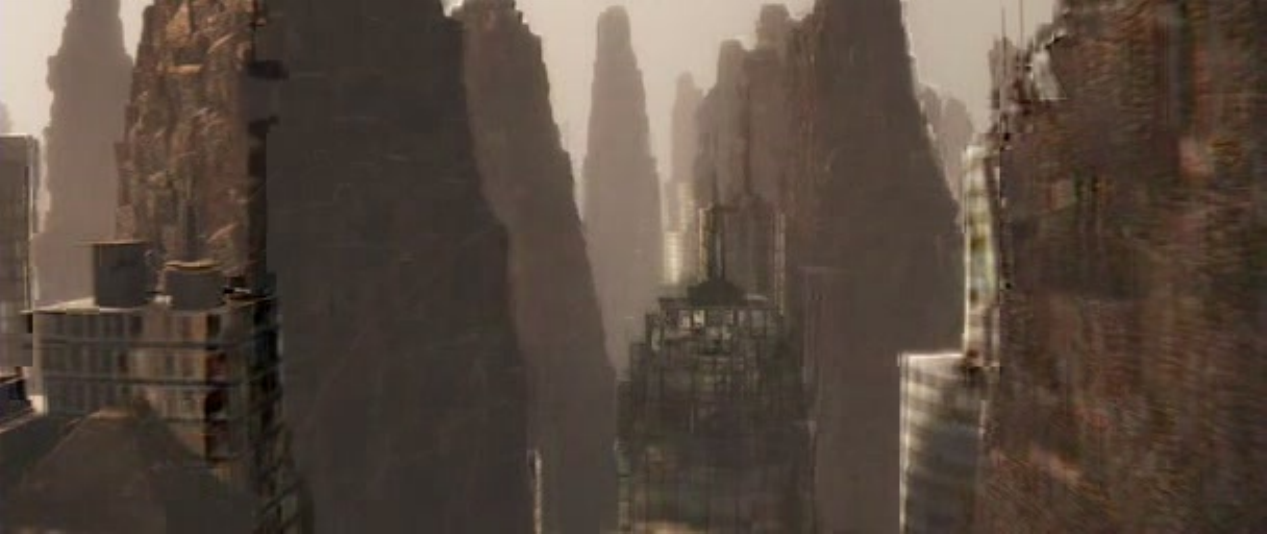} &
\includegraphics[width=0.46\textwidth]{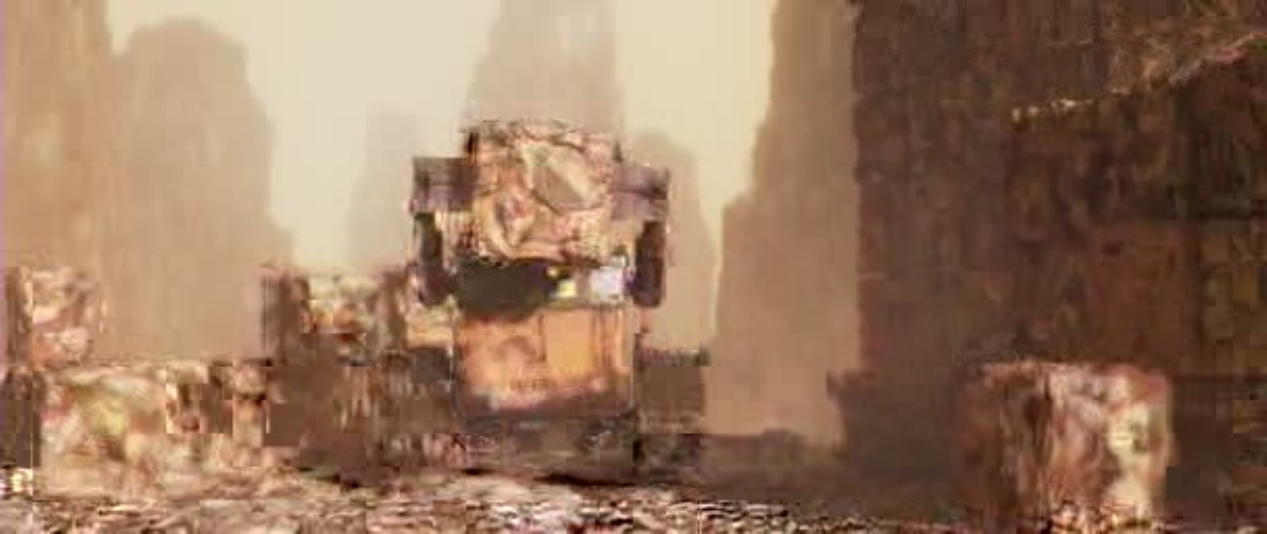} & \textbf{\small (1)}\\
\includegraphics[width=0.46\textwidth]{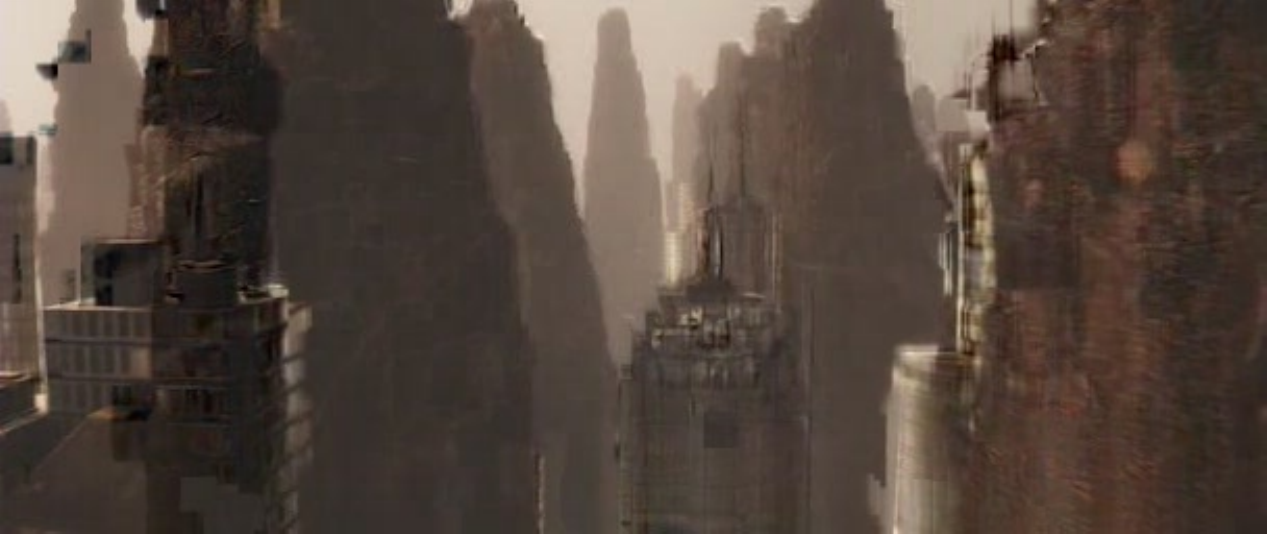} &
\includegraphics[width=0.46\textwidth]{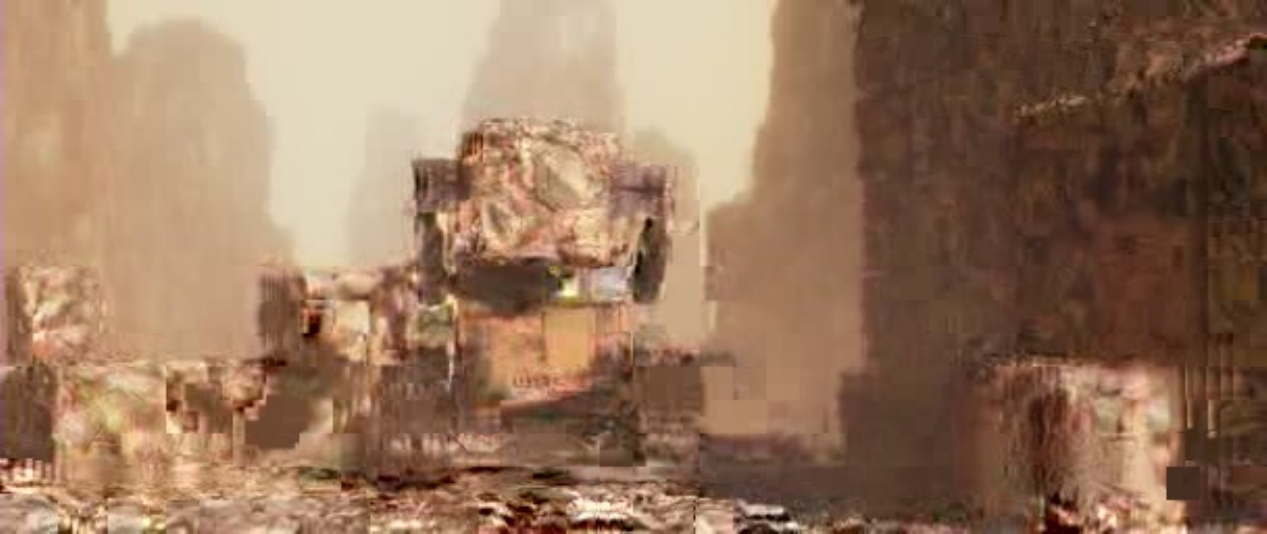} & \textbf{\small (2)}\\
\hspace{0.23\textwidth}\textbf{\small (c)} & \hspace{0.23\textwidth}\textbf{\small (d)} \\
\includegraphics[width=0.46\textwidth]{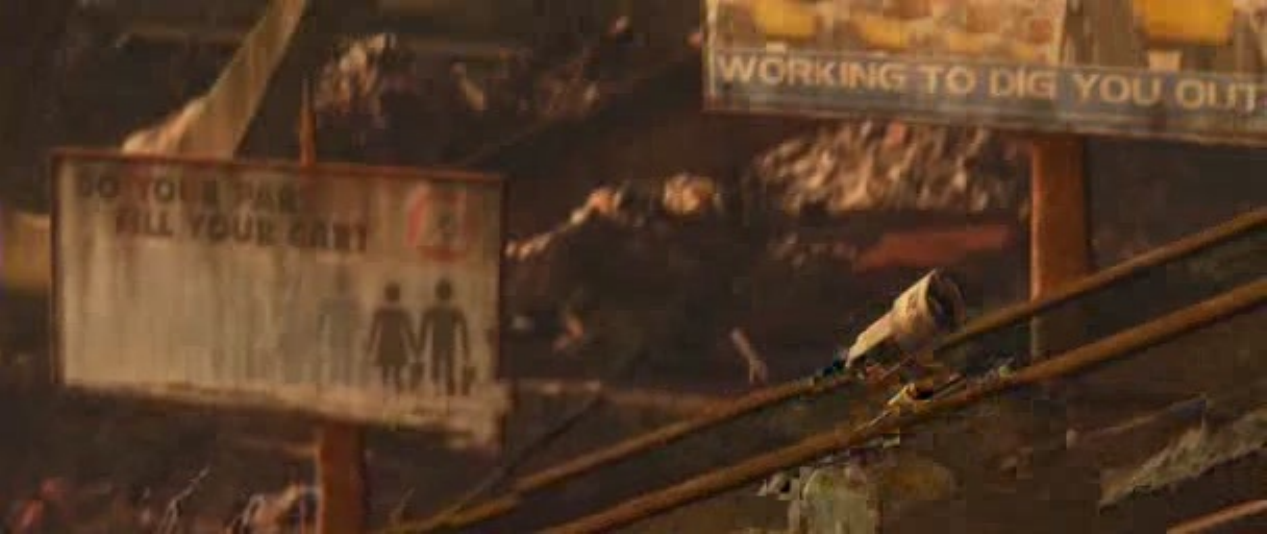} &
\includegraphics[width=0.46\textwidth]{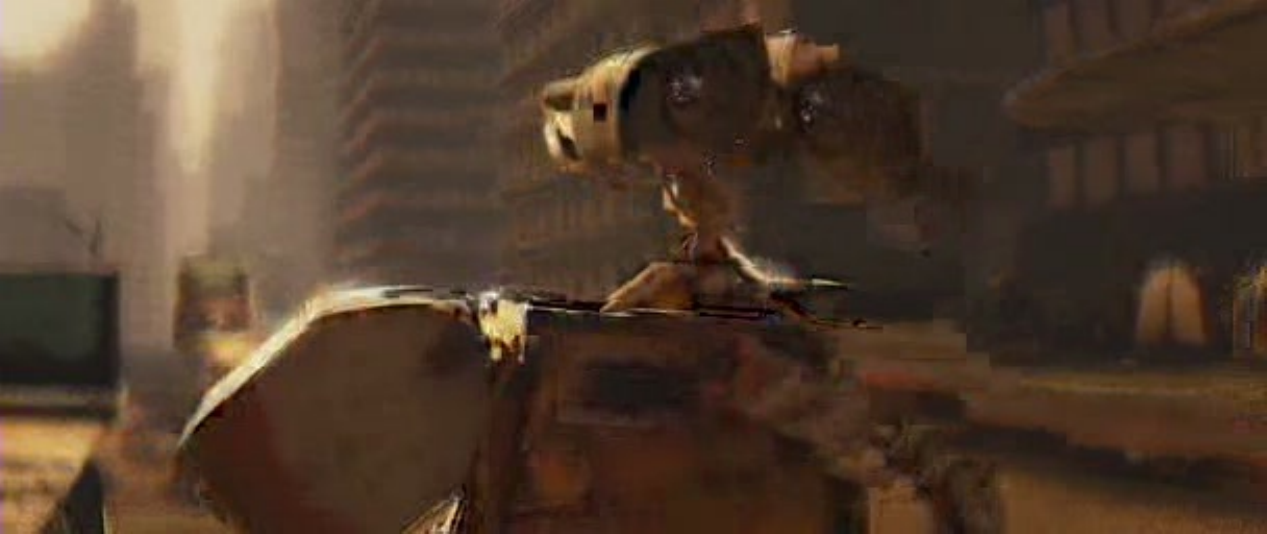} & \textbf{\small (1)}\\
\includegraphics[width=0.46\textwidth]{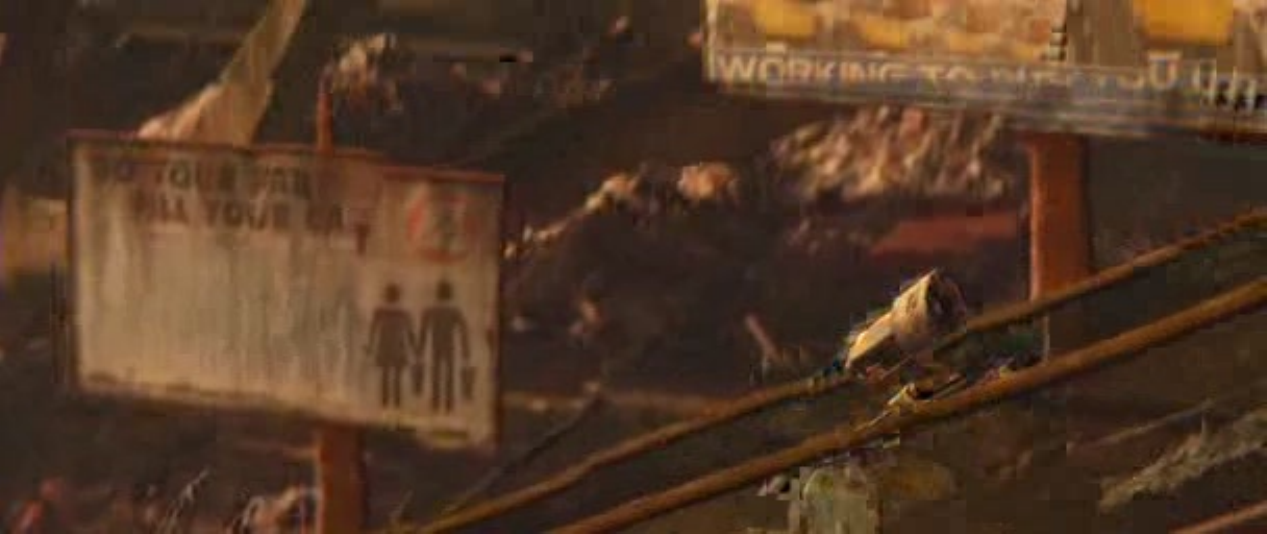} &
\includegraphics[width=0.46\textwidth]{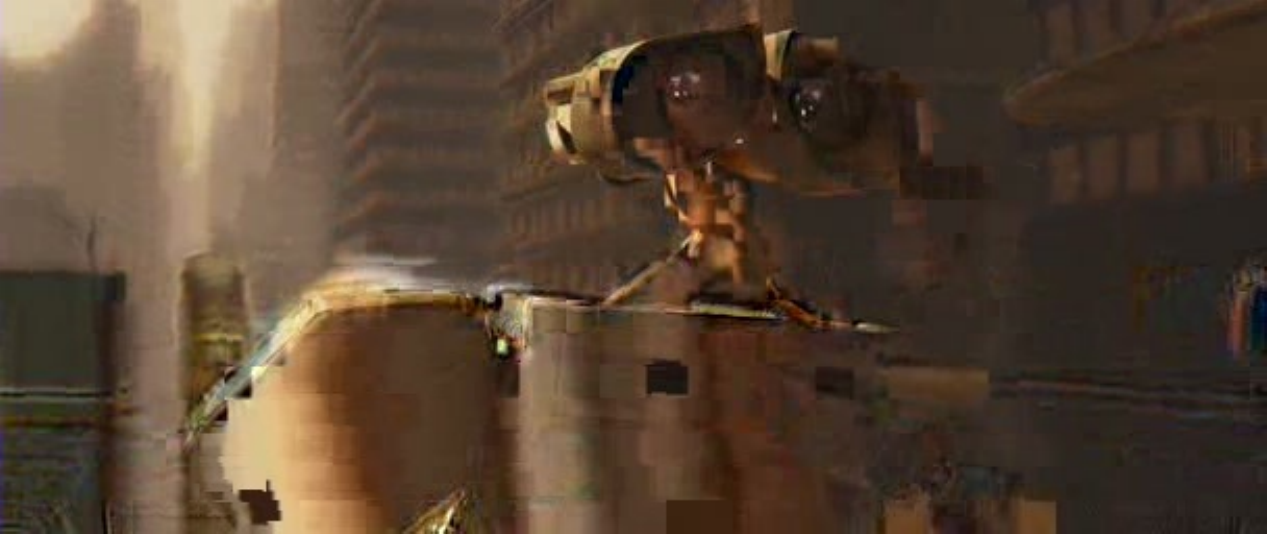} & \textbf{\small (2)}\\
\end{tabular}
\caption{Screenshots for different server setups: IEC without subchannels (1) and without FEC (2) ($M=7$, $R=1$, $p_e=0.1$).}
\label{screenshots1}
\end{figure*}

Below the admissible packet loss probability IEC-S, like every other system, behaves poorly in terms of decoding efficiency as shown in Figure \ref{pnb-pr}. However, as long as IEC-S decodes some packets, decoding efficiency will improve in every time slot. Thus, even with high packet loss rates, it must be possible to provide good playback quality by waiting for extra time slots until a client stores enough implicit redundancy to improve its packet loss protection over that of the network's.

\subsection{IEC-S Computational Complexity}

In terms of the computational complexity of IEC-S, the original IEC system has a computational complexity bounded by $\mathrm{O}(M^2)$ \cite{mtap1}, which is admissible considering that $M$ is generally a low number. Moreover, the IEC system with redundancy has a complexity bounded by $\mathrm{O}(I\cdot M) = \mathrm{O}(M^2 + R\cdot M) \leq \mathrm{O}(I^2)$. Thus, computational complexity depends on the number of channels, $I$. Consequently, IEC-S has a computational complexity bounded by $\mathrm{O}(\lambda^2 \cdot I \cdot M) = \mathrm{O}(\lambda^2 (M^2 + R \cdot M)) \leq \mathrm{O}(\lambda^2 \cdot I^2)$. 

It can be argued that if $\lambda$ is too high, it can lead to infeasible computational complexity. Obviously, $\lambda$ cannot experience unlimited growth, but our experiments show that relatively low values of $\lambda$ provide good performance, simultaneously producing computationally feasible coding matrices that any modern system can deal with. Thus, if the number of subchannels is moderate, the increase in computational complexity should be admissible for any commercially available playing device. As an example, in our tests (see Section \ref{num}), performed in commercial tablets and laptops, values of $\lambda=6$ or $\lambda=12$ (coding matrices of size $48\times 48$ or $96\times 96$, respectively) already showed very good performance.

\section{Test results}\label{num}

\subsection{Packet loss recovery}

The IEC server and client architectures have been presented in a previous paper \cite{mtap2}. Obviously, subchannels modify the architecture, but the changes are minor since, from the point of view of the implementation, a matrix of size $\lambda (M+R) \times \lambda M$ is treated as an $(M+R) \times M$ matrix. For this reason, in this paper we have not developed new architectures for the client or server. The work in \cite{mtap2} can be consulted for further details on IEC server and client architectures. 

The goal of our tests was to study the impact of subchannels on overall IEC performance. Figure \ref{conred} depicts diverse results for a setup with $M=7$ and $R=1$ ranging from 1 to 16 subchannels and for a packet loss probability $p_e=0.1$. For the first slot, this setup provides theoretical protection for an error probability $R/(M+R) = 0.125$ and, as expected from the theory, the tests show that increasing the number of subchannels improves packet loss recovery. However, this improvement is only noticeable at the beginning, which is also consistent with our theory and, beyond a certain number of subchannels, the benefit becomes almost negligible (see Figure \ref{pnb-pr}). Although it might seem that improving decoding efficiency in just a few slots is not very relevant, it is also true that due to the nature of the IEC schemes, most decoding problems appear during the first slots, as will be discussed later in this Section. Again, as predicted by the theory, Figure \ref{sinred} illustrates that the inclusion of subchannels requires the addition of redundancy to achieve full packet loss recovery. As has been mentioned in previous sections, subchannels do not provide higher packet loss protection but improve decoding efficiency, provided there is enough redundancy. However, improving decoding efficiency during the first time slots also implies the availability of more implicit redundancy during the subsequent time slots. Thus, subchannels improve decoding efficiency during the first time slots but, as a collateral effect, also improve packet loss protection in the mid and long-term thanks to an increase in implicit redundancy.

In Section \ref{sub},  we studied the effects of a packet loss probability beyond redundancy protection. We observed that under this circumstance the capacity to correct errors tends to zero. However, as can be observed in Figure \ref{alto}, IEC-S can overcome this problem, even for very high packet loss probabilities. The higher the packet loss probability, the longer IEC-S will require to achieve full packet recovery but, thanks to the utilization of previously downloaded packets as implicit redundancy, it is always possible to avoid null packet recovery, as might happen in other nVoD schemes, such as the work in \cite{jenkac} and \cite{Jenkac2006}. It is also important to stress that in the examples in Figure \ref{alto}, the redundancy channel only provides protection of packet losses of up to 12.5\%, which in most cases is less than the packet loss probability ($p_e$). Even for these cases, IEC-S can achieve full packet recovery after an amount of time that depends on the packet loss probability. Once the implicit redundancy protection exceeds the packet loss rate, the system behaves better than the original IEC scheme due to the utilization of subchannels (see Figure \ref{conred}).

\begin{figure*}[tbp!]
\centering
\setlength{\tabcolsep}{2pt}
\begin{tabular}{m{0.46\textwidth}m{0.46\textwidth}m{0.05\textwidth}}
\hspace{0.23\textwidth}\textbf{\small (a)} & \hspace{0.23\textwidth}\textbf{\small (b)} &\\
\includegraphics[width=0.46\textwidth]{0imagen100_48} &
\includegraphics[width=0.46\textwidth]{0imagen140_534} & \textbf{\small (1)}\\
\includegraphics[width=0.46\textwidth]{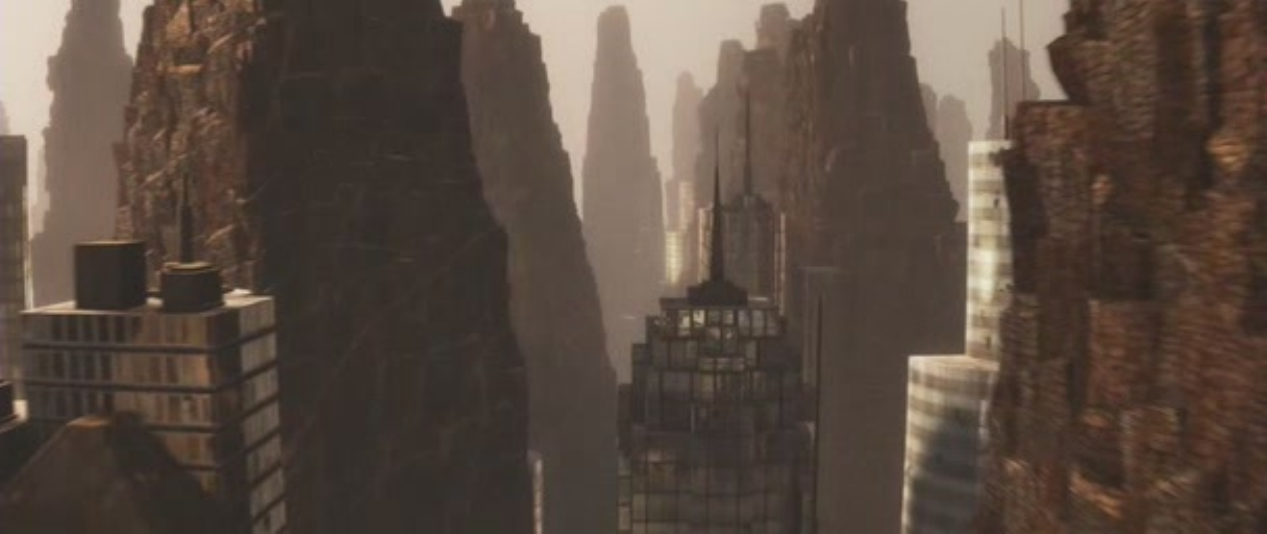} &
\includegraphics[width=0.46\textwidth]{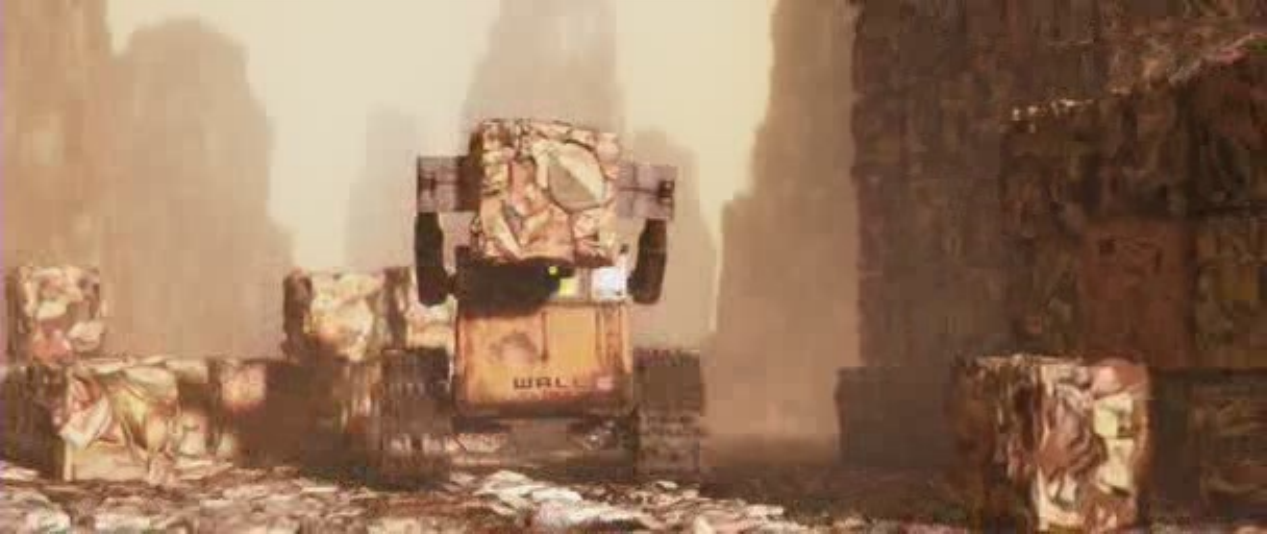} & \textbf{\small (2)}\\
\hspace{0.23\textwidth}\textbf{\small (c)} & \hspace{0.23\textwidth}\textbf{\small (d)} &\\\includegraphics[width=0.46\textwidth]{0imagen230_197} &
\includegraphics[width=0.46\textwidth]{0imagen230_324} & \textbf{\small (1)}\\
\includegraphics[width=0.46\textwidth]{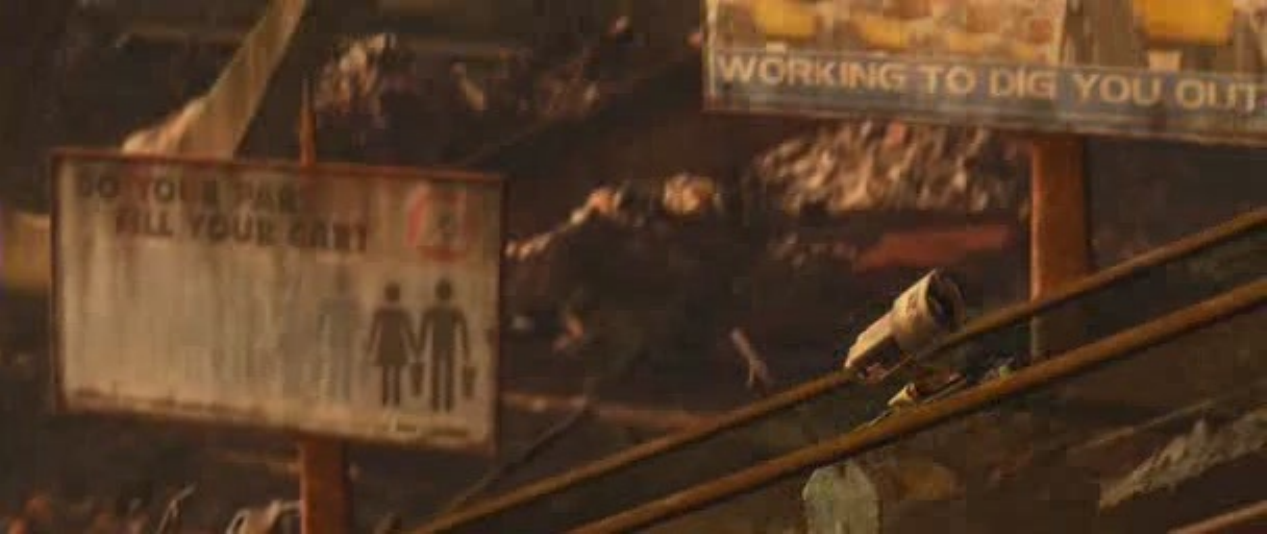} &
\includegraphics[width=0.46\textwidth]{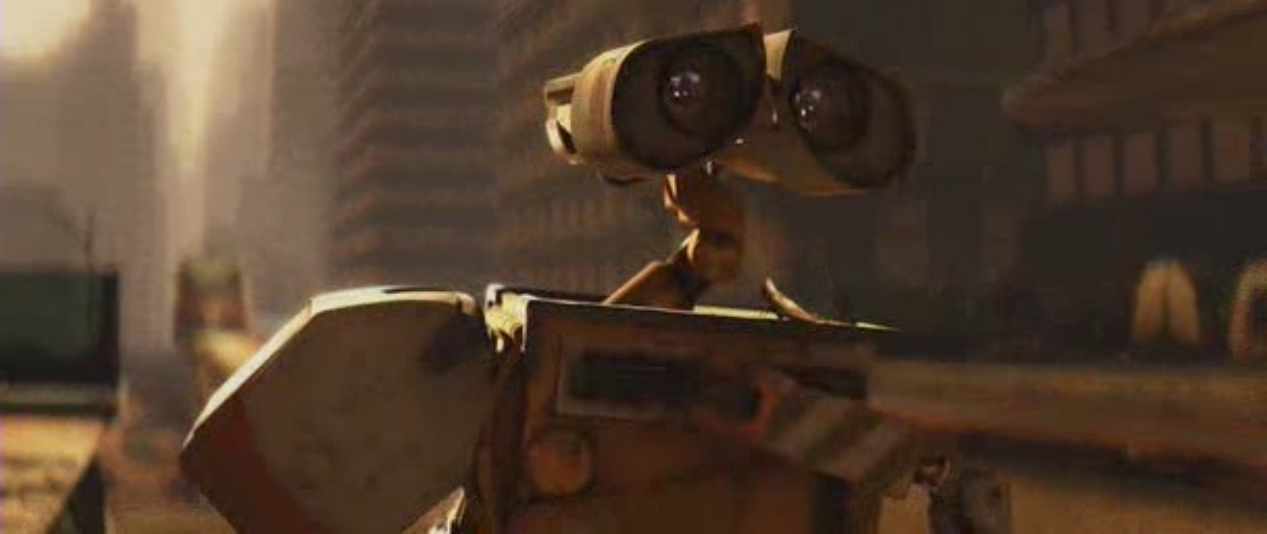} & \textbf{\small (2)}\\
\end{tabular}
\caption{Screenshots for different server setups: IEC without subchannels (1) and IEC for 4 subchannels (2) ($M=7$, $R=1$, $p_e=0.1$).}
\label{screenshots2}
\end{figure*}

\begin{figure}[tbp!]
\centering
\setlength{\tabcolsep}{2pt}
\begin{tabular}{m{0.46\columnwidth}m{0.46\columnwidth}m{0.05\columnwidth}}
\hspace{0.23\columnwidth}\textbf{\small (a)} & \hspace{0.23\columnwidth}\textbf{\small (b)} &\\
\includegraphics[width=0.46\columnwidth]{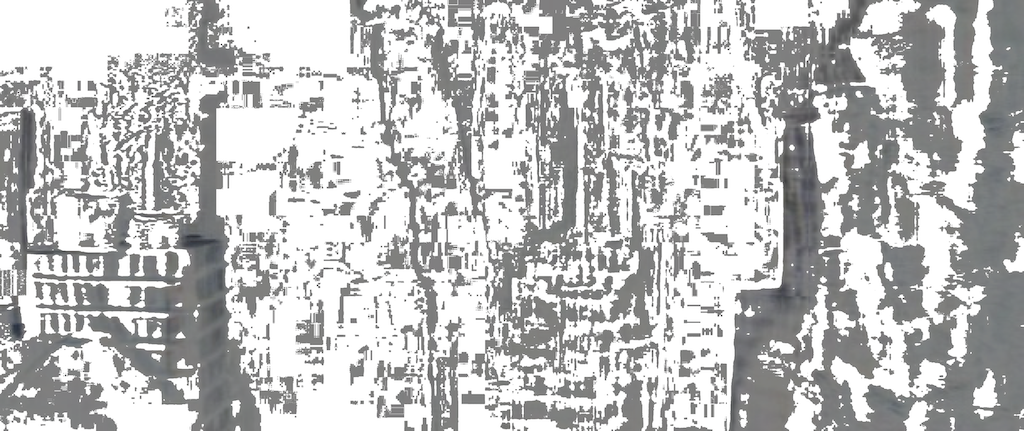} &
\includegraphics[width=0.46\columnwidth]{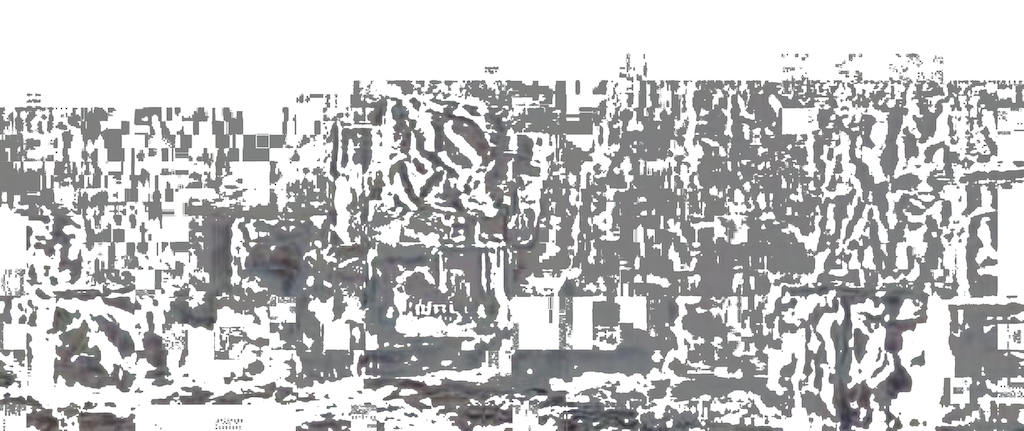} & \textbf{\small (1)}\\
\includegraphics[width=0.46\columnwidth]{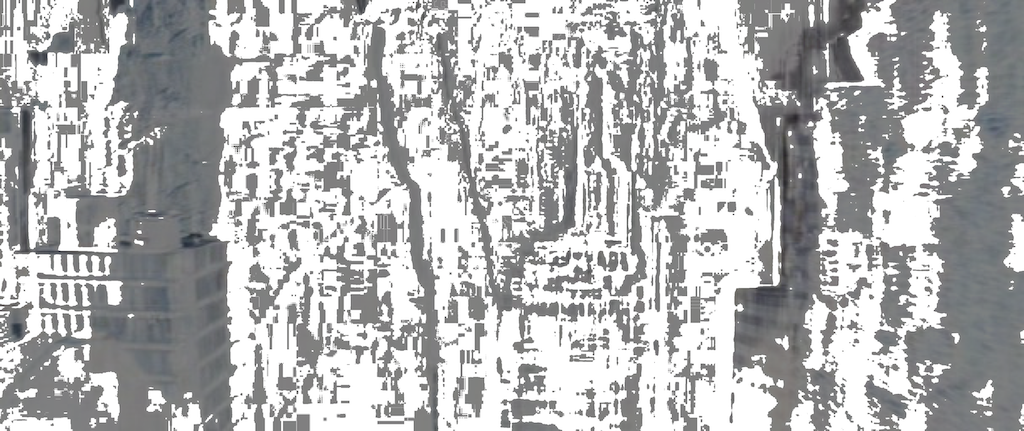} &
\includegraphics[width=0.46\columnwidth]{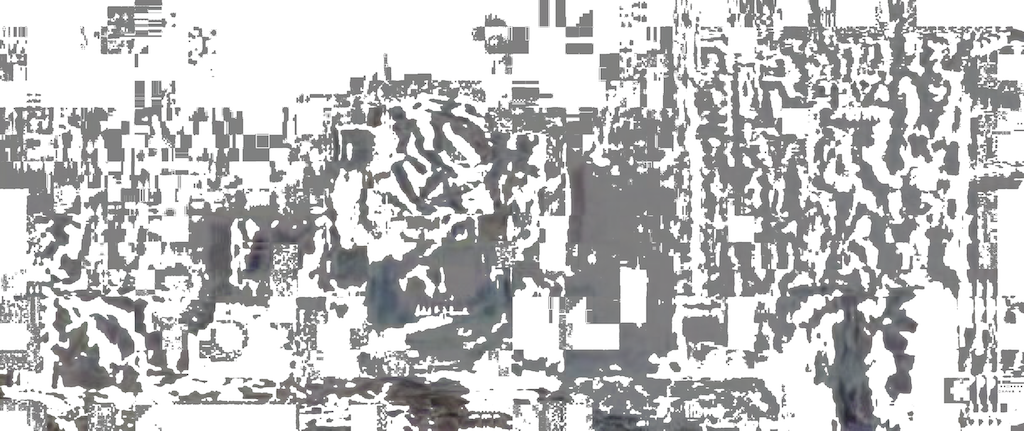} & \textbf{\small (2)}\\
\hspace{0.23\columnwidth}\textbf{\small (c)} & \hspace{0.23\columnwidth}\textbf{\small (d)} &\\
\includegraphics[width=0.46\columnwidth]{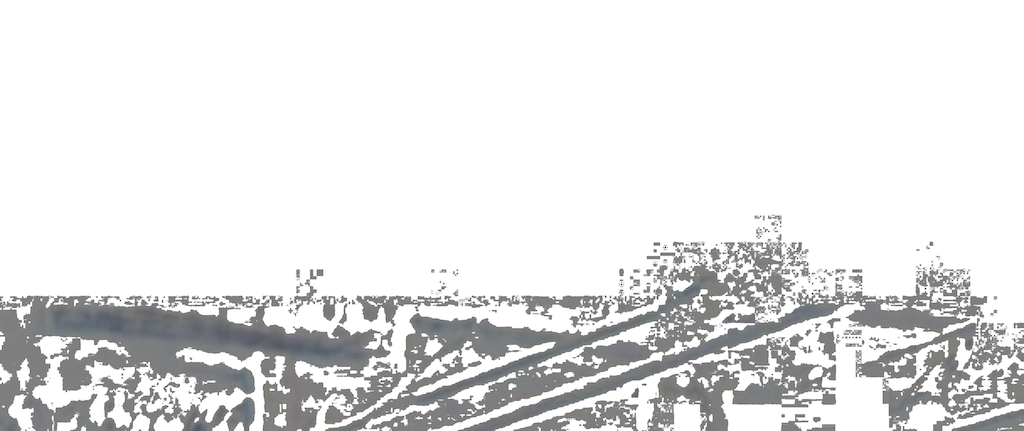} &
\includegraphics[width=0.46\columnwidth]{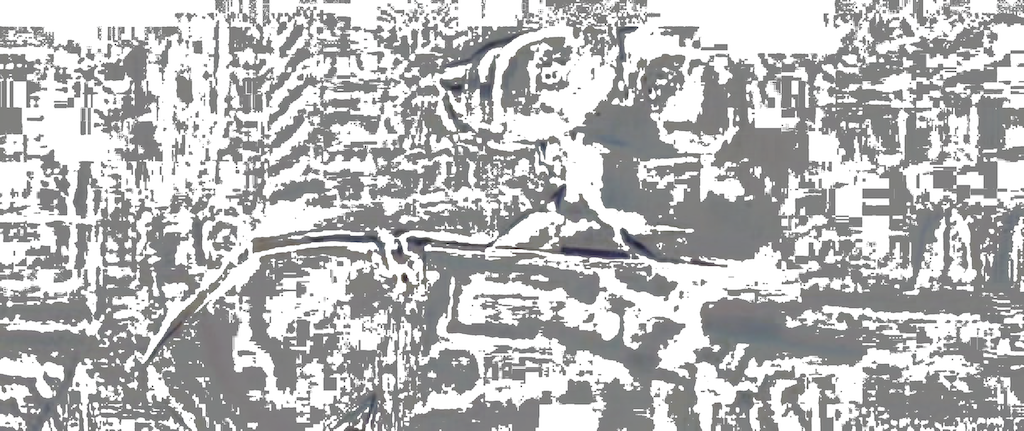} & \textbf{\small (1)}\\
\includegraphics[width=0.46\columnwidth]{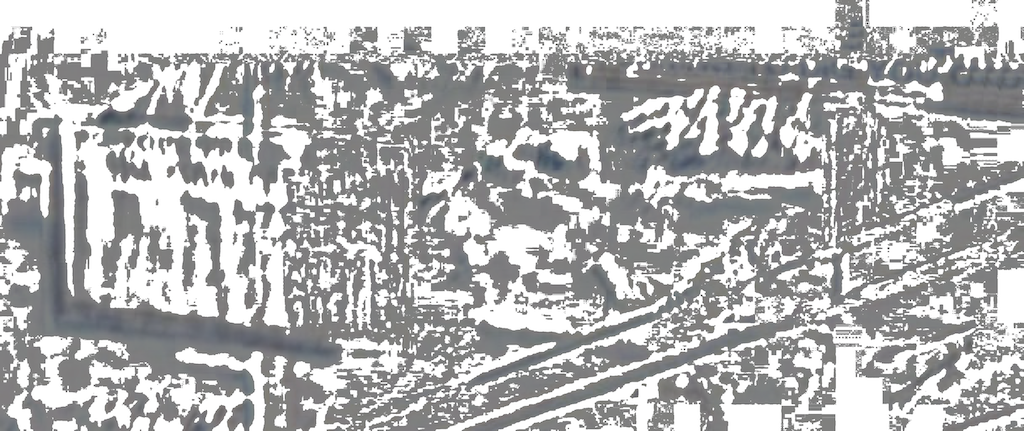} &
\includegraphics[width=0.46\columnwidth]{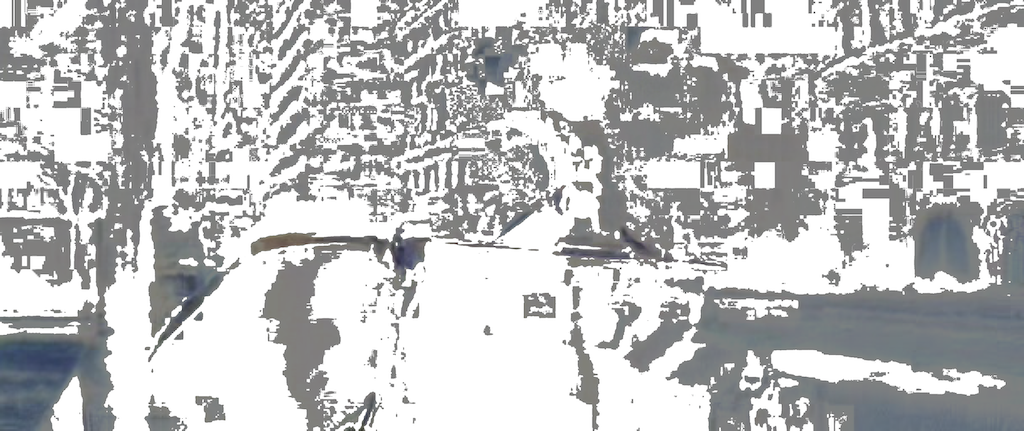} & \textbf{\small (2)}\\
\end{tabular}
\caption{Screenshots showing playback errors regarding IEC for 4 subchannels for different server setups: IEC without subchannels (1) and IEC without FEC (2) ($M=7$, $R=1$, $p_e=0.1$).}
\label{screenshots3}
\end{figure}

\subsection{Y-PSNR analysis}

Figure \ref{psnr} shows a Y-PSNR (Peak Signal-to-Noise Ratio of image luminance) analysis for different IEC-S setups. All tests were performed using the movie ``Wall-E'' (2008) and encoded with H.264 in HD with a data rate of \SI{2}{\mega\bit\per\second}. Y-PSNR is typically employed to assess video quality objectively without introducing significant errors \cite{bing2015next}. The first two parts of the graph correspond to the same scenario but for a different number of subchannels. As can be observed, if the system is appropriately dimensioned, a higher number of subchannels offers better subjective quality, although beyond a certain number of subchannels its benefit becomes almost unnoticeable. As previously mentioned, most decoding problems appear during the first time slots and, as a consequence, all improvements to the IEC scheme that are relevant for the first time slots can be considered to be significant. At the bottom of Figure \ref{psnr} we show an example for an under-dimensioned setup with a packet loss probability $p_e=0.25$. Although the subjective quality is poor at the beginning, it converges to that of the other examples before reaching a playback time of 10\%, which demonstrates the robustness of the IEC-S scheme under adverse conditions.

Finally, we illustrate some sample screenshots from the movie ``Wall-E'' in Figures \ref{screenshots1} and \ref{screenshots2}. From these figures, with a packet loss probability of $p_e=0.1$, it can be observed that $\lambda=4$ subchannels are enough to achieve good results compared to those provided by IEC without subchannels and IEC without FEC (Figure \ref{screenshots2}, (2), (a)-(c)). Figure \ref{screenshots3} shows the image errors regarding the playback of the IEC scheme using 4 subchannels. As can be expected, the IEC setup with $M=7$ and $R=1$ behaves better than IEC without subchannels. Nevertheless, these screenshots show that only 4 subchannels are enough to significantly improve playback quality. This is consistent with all the issues previously discussed in this paper.

\section{Conclusions and future lines}\label{conclusions}

This paper presents the IEC-S scheme. It extends the IEC scheme in \cite{mtap1} and \cite{mtap2} by introducing subchannels. Specifically, a channel is divided into $\lambda$ subchannels, each using $1/\lambda$ of the channel bandwidth. This allows for better decoding efficiency and better objective playback quality at the cost of a moderate increase in computational complexity.

Subchannels alone do not increase error protection, but rather error recovery efficiency. Intuitively, using subchannels increases tcoding and decoding matrix sizes. This means that the same number of packet losses, which in the original IEC scheme were spread over a few channels, are distributed over many sub-channels in IEC-S, which makes it more robust thanks to the fact that the decoding system can operate more efficiently. As long as the decoder is efficient when the number of packets is large enough, it is preferable that the number of packet losses be bounded rather than highly variable when the number of channels is low.

Despite its benefits, the use of subchannels does not improve error recovery efficiency if the system is not well dimensioned. This happens when the packet loss probability ($p_e$) is higher than the maximum admissible packet loss probability ($p_{n,b,R,\lambda}$) of the nVoD scheme. However, although for other nVoD schemes this might result in very poor performance, IEC-S maximum admissible packet loss probability increases with each time slot and, at some point, exceeds this probability ($p_{n,b,R,\lambda} > p_e$). When this happens, the subchannel scheme outperforms the original IEC mechanism. 

The subchannel scheme does not require modifications to the original architecture. From the point of view of implementation, a subchannel is equivalent to an original channel. For this reason, the subchannel technique is  easily integrated while providing important benefits in terms of playback quality.

The introduction of subchannels significantly improves IEC performance and playback quality. The current IEC scheme provides similar packet loss protection to all channels, although some segments are more sensitive than others (lower numbered segments are more sensitive to errors). One of the future lines that is under development is the introduction of asymmetric packet loss protection. The goal is to provide higher packet loss protection to those segments which are more sensitive to packet losses. 

IEC-S and all previous variants do not focus on bandwidth optimization; i.e. it uses a fixed number of channels and the more channels there are, the shorter the playback delay will be. This design is efficient in the sense that it supports a virtually infinite number of clients. However, if the number of clients is low (lower than the number of channels), unicast transmissions would be more efficient in terms of bandwidth. Moreover, IEC-S implicit redundancy implies that at a certain point, the client's maximum admissible packet loss probability is much higher than the network packet loss probability. In fact, it would not be necessary to download all the segments of a given time slot but a subset of them to allow full decoding. Another future line is to design an IEC scheme that also optimizes bandwidth utilization by allowing clients to select which segments to download for every time slot.

Finally, the IEC scheme is designed for video distribution and achieves its potential in limited bandwidth error prone environments. Albeit video distribution being its main application, IEC could be applied to other scenarios, such as software updates in large wireless sensor networks or Internet of Things applications. In general, IEC might be very helpful whenever it is required to distribute digital contents to large audiences.

\section*{Acknowledgements}	
This research has been supported by the project AIM, ref. TEC2016-76465-C2-1-R (AEI/FEDER, UE).

\bibliographystyle{IEEEtran}
\bibliography{iec}

\begin{thebibliography}{10}
\providecommand{\url}[1]{#1}
\csname url@samestyle\endcsname
\providecommand{\newblock}{\relax}
\providecommand{\bibinfo}[2]{#2}
\providecommand{\BIBentrySTDinterwordspacing}{\spaceskip=0pt\relax}
\providecommand{\BIBentryALTinterwordstretchfactor}{4}
\providecommand{\BIBentryALTinterwordspacing}{\spaceskip=\fontdimen2\font plus
\BIBentryALTinterwordstretchfactor\fontdimen3\font minus
  \fontdimen4\font\relax}
\providecommand{\BIBforeignlanguage}[2]{{%
\expandafter\ifx\csname l@#1\endcsname\relax
\typeout{** WARNING: IEEEtran.bst: No hyphenation pattern has been}%
\typeout{** loaded for the language `#1'. Using the pattern for}%
\typeout{** the default language instead.}%
\else
\language=\csname l@#1\endcsname
\fi
#2}}
\providecommand{\BIBdecl}{\relax}
\BIBdecl

\bibitem{6587820}
Y.~Chen, B.~Zhang, Y.~Liu, and W.~Zhu, ``{Measurement and Modeling of Video
  Watching Time in a Large-Scale Internet Video-on-Demand System},'' \emph{IEEE
  Transactions on Multimedia}, vol.~15, no.~8, pp. 2087--2098, Dec 2013.

\bibitem{7128393}
Y.~Zhou, L.~Chen, C.~Yang, and D.~M. Chiu, ``{Video Popularity Dynamics and Its
  Implication for Replication},'' \emph{IEEE Transactions on Multimedia},
  vol.~17, no.~8, pp. 1273--1285, Aug 2015.

\bibitem{7155559}
H.~Hu, Y.~Wen, T.~Chua, J.~Huang, W.~Zhu, and X.~Li, ``{Joint Content
  Replication and Request Routing for Social Video Distribution Over Cloud CDN:
  A Community Clustering Method},'' \emph{IEEE Transactions on Circuits and
  Systems for Video Technology}, vol.~26, no.~7, pp. 1320--1333, July 2016.

\bibitem{mtap1}
\BIBentryALTinterwordspacing
F.~Gonz\'alez-Casta\~no, R.~Asorey-Cacheda, H.~Cerezo-Costas,
  J.~Burguillo-Rial, and F.~Gil-Casti\~neira, ``{A zero-overhead
  error-correcting nVoD schema},'' \emph{Multimedia Tools and Applications},
  vol.~48, pp. 291--312, 2010, 10.1007/s11042-009-0331-7. [Online]. Available:
  \url{http://dx.doi.org/10.1007/s11042-009-0331-7}
\BIBentrySTDinterwordspacing

\bibitem{mtap2}
\BIBentryALTinterwordspacing
R.~Asorey-Cacheda, B.~Pedrero-L\'opez, and F.~Gonz\'alez-Casta\~no,
  ``\BIBforeignlanguage{English}{{Efficient implementation of the implicit
  error correction nVoD schema}},''
  \emph{\BIBforeignlanguage{English}{Multimedia Tools and Applications}},
  vol.~64, no.~3, pp. 627--647, 2013. [Online]. Available:
  \url{http://dx.doi.org/10.1007/s11042-011-0966-z}
\BIBentrySTDinterwordspacing

\bibitem{pb2X}
\BIBentryALTinterwordspacing
K.~A. Hua, ``{Online Video Delivery: Past, Present, and Future},'' \emph{ACM
  Trans. Multimedia Comput. Commun. Appl.}, vol.~9, no.~1s, pp. 39:1--39:4, 10
  2013. [Online]. Available: \url{http://doi.acm.org/10.1145/2502435}
\BIBentrySTDinterwordspacing

\bibitem{harmonic}
\BIBentryALTinterwordspacing
Y.~Sun and T.~Kameda, ``{Harmonic Block Windows Scheduling Through Harmonic
  Windows Scheduling},'' in \emph{Advances in Multimedia Information Systems},
  ser. Lecture Notes in Computer Science, K.~Candan and A.~Celentano,
  Eds.\hskip 1em plus 0.5em minus 0.4em\relax Springer Berlin / Heidelberg,
  2005, vol. 3665, pp. 190--206, 10.1007/11551898\_17. [Online]. Available:
  \url{http://dx.doi.org/10.1007/11551898\_17}
\BIBentrySTDinterwordspacing

\bibitem{7782746}
X.~Tian, C.~Zhao, H.~Liu, and J.~Xu, ``{Video On-Demand Service via Wireless
  Broadcasting},'' \emph{IEEE Transactions on Mobile Computing}, vol.~16,
  no.~10, pp. 2970--2982, 10 2017.

\bibitem{Altaf2017}
\BIBentryALTinterwordspacing
M.~Altaf, F.~A. Khan, N.~Qadri, M.~Ghanbari, and S.~E. Dudley, ``{Adaptive
  robust video broadcast via satellite},'' \emph{Multimedia Tools and
  Applications}, vol.~76, no.~6, pp. 7785--7801, 3 2017. [Online]. Available:
  \url{https://doi.org/10.1007/s11042-016-3426-y}
\BIBentrySTDinterwordspacing

\bibitem{290771X}
K.~Nayfeh and N.~Sarhan, ``{Design and analysis of scalable and interactive
  near video-on-demand systems},'' in \emph{Multimedia and Expo (ICME), 2013
  IEEE International Conference on}, July 2013, pp. 1--6.

\bibitem{7552967}
K.~K. Nayfeh and N.~J. Sarhan, ``{Client-side cache management for scalable and
  interactive video streaming},'' in \emph{2016 IEEE International Conference
  on Multimedia and Expo (ICME)}, July 2016, pp. 1--6.

\bibitem{8221104}
J.~Yuan, X.~Wang, and L.~Xiao, ``{Hybrid Video Transmission Scheme for
  Minimizing Maximum Waiting Time in Video-on-Demand (VOD) System},'' in
  \emph{2017 2nd International Conference on Multimedia and Image Processing
  (ICMIP)}, 3 2017, pp. 225--229.

\bibitem{eager99bandwidthX}
K.~Nayfeh and N.~Sarhan, ``{A Scalable Solution for Interactive Near
  Video-on-Demand Systems},'' \emph{Circuits and Systems for Video Technology,
  IEEE Transactions on}, vol.~PP, no.~99, pp. 1--1, 2015.

\bibitem{10.1007/978-981-10-7605-3_124}
S.~Kim and Y.~Won, ``{Frame Rate Control Buffer Management Technique for
  High-Quality Real-Time Video Conferencing System},'' in \emph{Advances in
  Computer Science and Ubiquitous Computing}, J.~J. Park, V.~Loia, G.~Yi, and
  Y.~Sung, Eds.\hskip 1em plus 0.5em minus 0.4em\relax Singapore: Springer
  Singapore, 2018, pp. 777--783.

\bibitem{Abozeid:2017:SVS:3093241.3093287}
\BIBentryALTinterwordspacing
A.~Abozeid, H.~Farouk, and K.~ElDahshan, ``{Scalable Video Summarization: A
  Comparative Study},'' in \emph{Proceedings of the International Conference on
  Compute and Data Analysis}, ser. ICCDA '17.\hskip 1em plus 0.5em minus
  0.4em\relax New York, NY, USA: ACM, 2017, pp. 215--219. [Online]. Available:
  \url{http://doi.acm.org/10.1145/3093241.3093287}
\BIBentrySTDinterwordspacing

\bibitem{940034X}
K.~Anang, P.~Rpajic, T.~Eneh, and Y.~Nijsure, ``{Minimum Cell Size for
  Information Capacity Increase in Cellular Wireless Network},'' in
  \emph{Vehicular Technology Conference (VTC Spring), 2011 IEEE 73rd}, May
  2011, pp. 1--6.

\bibitem{sp}
D.~M. Dobkin, \emph{{RF Engineering for Wireless Networks: Hardware, Antennas
  and Propagation}}.\hskip 1em plus 0.5em minus 0.4em\relax Newnes, 2004.

\bibitem{er10}
J.~G. Andrews, A.~Ghosh, and R.~Muhamed, \emph{{Fundamentals of WiMAX:
  Understanding Broadband Wireless Networking (Prentice Hall Communications
  Engineering and Emerging Technologies Series)}}.\hskip 1em plus 0.5em minus
  0.4em\relax Upper Saddle River, NJ, USA: Prentice Hall PTR, 2007.

\bibitem{raso08}
R.~Asorey-Cacheda and F.~Gonz\'alez-Casta\~no, ``{A Multicast nVoD Schema with
  Zero-Overhead Implicit Error Correction},'' in \emph{IEEE International
  Conference on Communications, 2008. ICC '08.}, 5 2008, pp. 2017--2020.

\bibitem{5425298}
R.~Asorey-Cacheda, H.~Huang, F.~Gonz\'alez-Casta\~no, E.~Johnson,
  C.~L\'opez-Bravo, and F.~Gil-Casti\~neira, ``{A Joint Interchannel and
  Network Coding Schema for nVoD Services over Wireless Mesh Networks},'' in
  \emph{Global Telecommunications Conference, 2009. GLOBECOM 2009. IEEE}, Nov
  2009, pp. 1--8.

\bibitem{jenkac}
H.~Jenka{\v{c}} and T.~Stockhammer, ``{Asynchronous Media Streaming Over
  Wireless Broadcast Channels},'' in \emph{IEEE International Conference on
  Multimedia and Expo, 2005. ICME 2005.}, July 2005, pp. 1318--1321.

\bibitem{Jenkac2006}
\BIBentryALTinterwordspacing
H.~Jenka{\v{c}}, T.~Stockhammer, W.~Xu, and W.~Abdel~Samad, ``{Efficient
  Video-on-Demand services over mobile datacast channels},'' \emph{Journal of
  Zhejiang University-SCIENCE A}, vol.~7, no.~5, pp. 873--884, 5 2006.
  [Online]. Available: \url{https://doi.org/10.1631/jzus.2006.A0873}
\BIBentrySTDinterwordspacing

\bibitem{10.9717/JMIS.2018.5.1.27}
\BIBentryALTinterwordspacing
C.~Stergiou, K.~E. Psannis, A.~P. Plageras, Y.~Ishibashi, and B.-G. Kim,
  ``{Algorithms for Efficient Digital Media Transmission over IoT and Cloud
  Networking},'' \emph{Journal of Multimedia Information System}, vol.~5,
  no.~1, pp. 27--34, 2018. [Online]. Available:
  \url{https://doi.org/10.9717/JMIS.2018.5.1.27}
\BIBentrySTDinterwordspacing

\bibitem{MEMOS2018619}
\BIBentryALTinterwordspacing
V.~A. Memos, K.~E. Psannis, Y.~Ishibashi, B.-G. Kim, and B.~Gupta, ``{An
  Efficient Algorithm for Media-based Surveillance System (EAMSuS) in IoT Smart
  City Framework},'' \emph{Future Generation Computer Systems}, vol.~83, pp.
  619 -- 628, 2018. [Online]. Available:
  \url{http://www.sciencedirect.com/science/article/pii/S0167739X17307707}
\BIBentrySTDinterwordspacing

\bibitem{STERGIOU2018964}
\BIBentryALTinterwordspacing
C.~Stergiou, K.~E. Psannis, B.-G. Kim, and B.~Gupta, ``{Secure integration of
  IoT and Cloud Computing},'' \emph{Future Generation Computer Systems},
  vol.~78, pp. 964 -- 975, 2018. [Online]. Available:
  \url{http://www.sciencedirect.com/science/article/pii/S0167739X1630694X}
\BIBentrySTDinterwordspacing

\bibitem{PLAGERAS2018349}
\BIBentryALTinterwordspacing
A.~P. Plageras, K.~E. Psannis, C.~Stergiou, H.~Wang, and B.~Gupta, ``{Efficient
  IoT-based sensor BIG Data collection–processing and analysis in smart
  buildings},'' \emph{Future Generation Computer Systems}, vol.~82, pp. 349 --
  357, 2018. [Online]. Available:
  \url{http://www.sciencedirect.com/science/article/pii/S0167739X17314127}
\BIBentrySTDinterwordspacing

\bibitem{Psannis2006}
\BIBentryALTinterwordspacing
K.~E. Psannis and Y.~Ishibashi, ``{Impact of Video Coding on Delay and Jitter
  in 3G Wireless Video Multicast Services},'' \emph{EURASIP Journal on Wireless
  Communications and Networking}, vol. 2006, no.~1, p. 024616, Jul 2006.
  [Online]. Available: \url{https://doi.org/10.1155/WCN/2006/24616}
\BIBentrySTDinterwordspacing

\bibitem{6416071}
E.~Magli, M.~Wang, P.~Frossard, and A.~Markopoulou, ``{Network Coding Meets
  Multimedia: A Review},'' \emph{IEEE Transactions on Multimedia}, vol.~15,
  no.~5, pp. 1195--1212, Aug 2013.

\bibitem{6600846}
D.~Vukobratović, C.~Khirallah, V.~Stanković, and J.~S. Thompson, ``{Random
  Network Coding for Multimedia Delivery Services in LTE/LTE-Advanced},''
  \emph{IEEE Transactions on Multimedia}, vol.~16, no.~1, pp. 277--282, Jan
  2014.

\bibitem{8013842}
X.~Xu, Y.~Zeng, Y.~L. Guan, and L.~Yuan, ``{Expanding-Window BATS Code for
  Scalable Video Multicasting Over Erasure Networks},'' \emph{IEEE Transactions
  on Multimedia}, vol.~20, no.~2, pp. 271--281, Feb 2018.

\bibitem{6410040}
X.~Wang, M.~Chen, T.~T. Kwon, L.~Yang, and V.~C.~M. Leung, ``{AMES-Cloud: A
  Framework of Adaptive Mobile Video Streaming and Efficient Social Video
  Sharing in the Clouds},'' \emph{IEEE Transactions on Multimedia}, vol.~15,
  no.~4, pp. 811--820, June 2013.

\bibitem{5418877}
R.~Asorey-Cacheda, H.~Cerezo-Costas, F.~J. González-Castaño, and F.~J.
  Gil-Castiñeira, ``{A zero-overhead implicit error correction nVoD schema for
  scalable video},'' in \emph{2010 Digest of Technical Papers International
  Conference on Consumer Electronics (ICCE)}, Jan 2010, pp. 353--354.

\bibitem{juhn97harmonicX}
C.~Jayasundara, M.~Zukerman, T.~Nirmalathas, E.~Wong, and C.~Ranaweera,
  ``{Improving Scalability of VoD Systems by Optimal Exploitation of Storage
  and Multicast},'' \emph{Circuits and Systems for Video Technology, IEEE
  Transactions on}, vol.~24, no.~3, pp. 489--503, March 2014.

\bibitem{juhn2X}
\BIBentryALTinterwordspacing
H.~Febiansyah and J.~Kwon, ``\BIBforeignlanguage{English}{{Dynamic
  proxy-assisted scalable broadcasting of videos for heterogeneous
  environments}},'' \emph{\BIBforeignlanguage{English}{Multimedia Tools and
  Applications}}, vol.~66, no.~3, pp. 517--543, 2013. [Online]. Available:
  \url{http://dx.doi.org/10.1007/s11042-012-1044-x}
\BIBentrySTDinterwordspacing

\bibitem{8012499}
H.~Feng, Z.~Chen, and H.~Liu, ``{Design and Optimization of VoD Schemes With
  Client Caching in Wireless Multicast Networks},'' \emph{IEEE Transactions on
  Vehicular Technology}, vol.~67, no.~1, pp. 765--780, Jan 2018.

\bibitem{592604}
M.~{Zorzi} and R.~R. {Rao}, ``{On the statistics of block errors in bursty
  channels},'' \emph{IEEE Transactions on Communications}, vol.~45, no.~6, pp.
  660--667, June 1997.

\bibitem{Blomer95anxor-based}
J.~Blömer, M.~Kalfane, R.~Karp, M.~Karpinski, M.~Luby, and D.~Zuckerman, ``{An
  XOR-Based Erasure-Resilient Coding Scheme},'' 1995.

\bibitem{bing2015next}
\BIBentryALTinterwordspacing
B.~Bing, \emph{{Assessing and Enhancing Video Quality}}.\hskip 1em plus 0.5em
  minus 0.4em\relax Wiley, 2015, ch.~5. [Online]. Available:
  \url{https://books.google.es/books?id=xQueCAAAQBAJ}
\BIBentrySTDinterwordspacing

\end{thebibliography}

\begin{IEEEbiography}[{\includegraphics[width=1in,height=1.25in,clip,keepaspectratio]{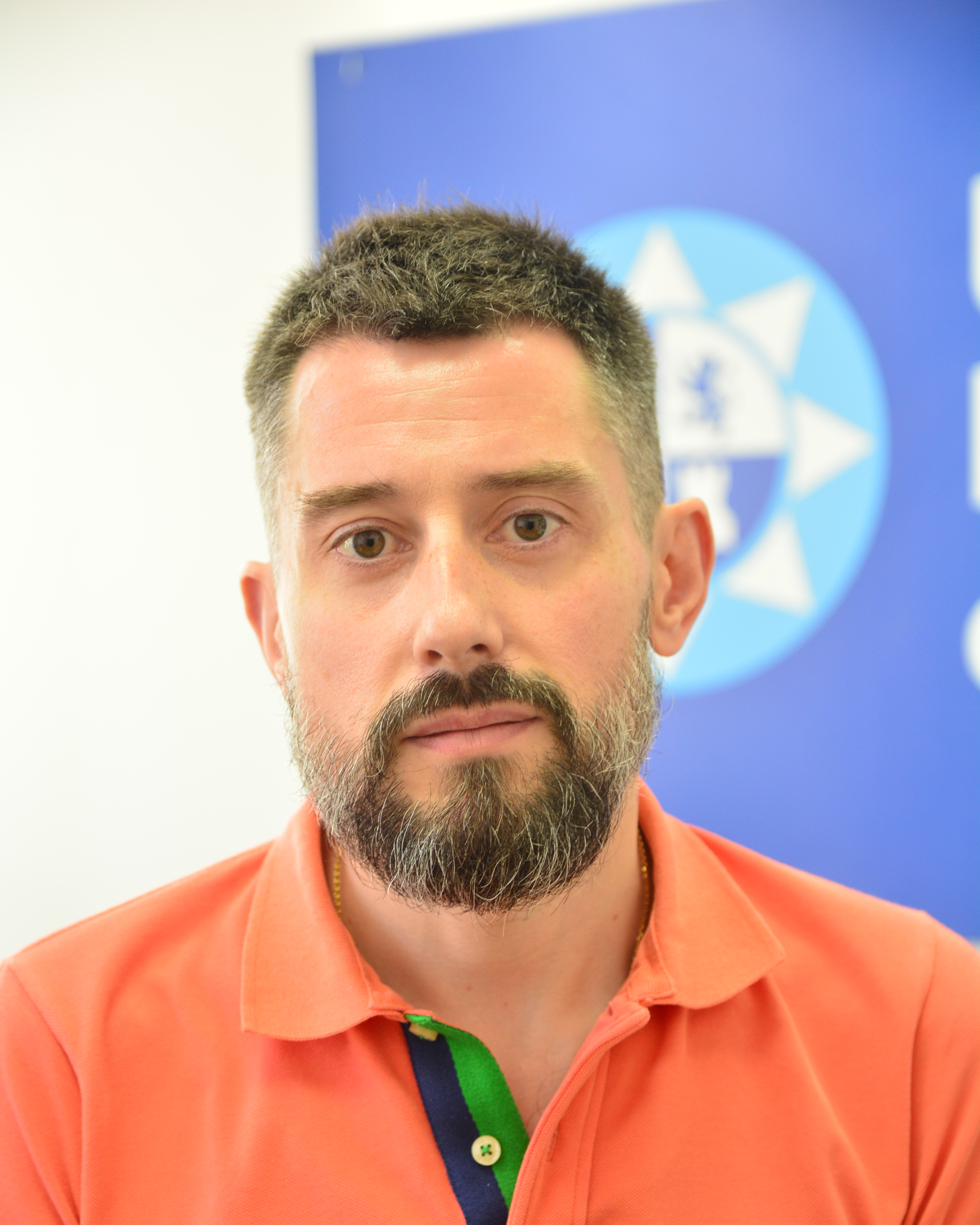}}]{Rafael Asorey-Cacheda}
received the M.Sc. degree in Telecommunication Engineering (major in Telematics and Best Master Thesis Award) and the Ph.D. Degree (cum laude and Best PhD Thesis Award) in Telecommunication Engineering from the Universidade de Vigo, Spain, in 2006 and 2009 respectively. He was a researcher with the Information Technologies Group, University of Vigo, Spain until 2009. Between 2008 and 2009 he was also R\&D Manager at Optare Solutions, a Spanish telecommunications company. Between 2009 and 2012 held an Ángeles Alvariño position, Xunta de Galicia, Spain. Between 2012 and 2018, he was an associate professor at the Centro Universitario de la Defensa en la Escuela Naval Militar, Universidade de Vigo. Currently, he is an associate professor at the Universidad Politécnica de Cartagena. He is author or co-author of more than 60 journal and conference papers mainly in the fields of switching, wireless networking and content distribution. He has been a visiting scholar at New Mexico State University, USA (2007-2011) and at Universidad Politécnica de Cartagena, Spain (2011, 2015). His interests include content distribution, high-performance switching, peer-to-peer networking, wireless networks, and nano-networks. He is currently granted by two national research (2006-2012, 2013-2018) and one national teaching (2010-2015) recognitions.
\end{IEEEbiography}

\begin{IEEEbiography}[{\includegraphics[width=1in,height=1.25in,clip,keepaspectratio]{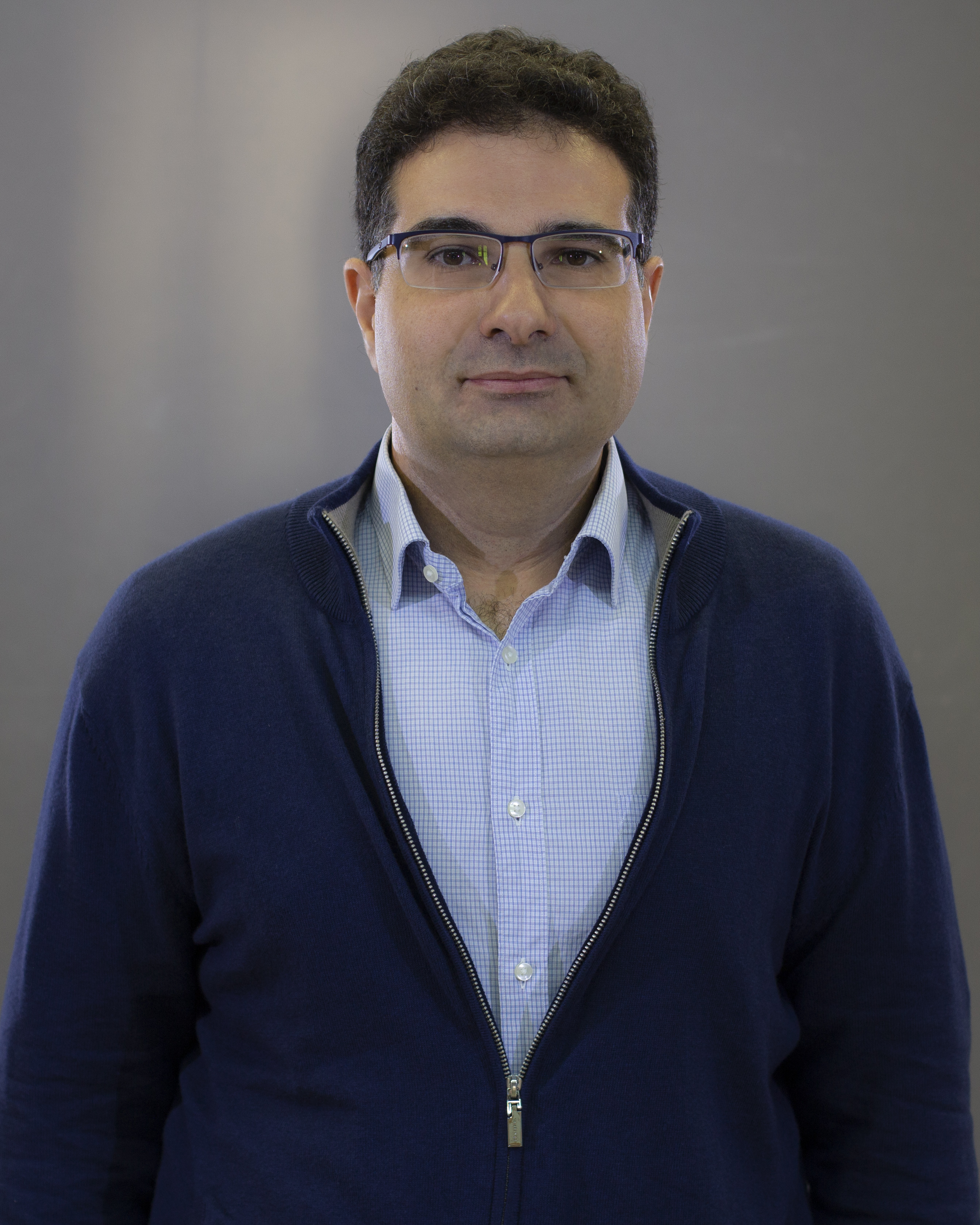}}]{Anonio-Javier Garcia-Sanchez}
received the Industrial Engineering degree (M.S. degree) in 2000 from the Technical University of Cartagena (UPCT), Spain. Since 2001, he has joined the to Information and Communication Technologies Department (DTIC), UPCT, obtaining the Ph.D. degree in 2005. Currently, he is Associate Professor at the UPCT and Head of the DTIC. He is a (co)author of more than 90 conference and journal papers, forty of them indexed in the Journal Citation Report (JCR). He has been the main head in several regional/national/EU research projects, and he is a reviewer of notable communication journals listed in the ISI-JCR, highlighting IEEE Internet of Things Journal, IEEE Journal on Selected Areas in Communications, IEEE Transactions on Industrial Electronics, Computer Networks, etc. Antonio has belonged to more than 30 conferences/workshops as TPC staff and to 8 as member of the organizing committee. His main research interests are in the areas of wireless sensor networks (WSNs) and Internet of Things, streaming services, performance evaluation of communication networks, smart data processing and nanocommunications. He is currently granted by two national research (2003-2010, 2011-2016) and three national teaching (2002-2006, 2007-2011, 2012-2016) recognitions. 
\end{IEEEbiography}

\begin{IEEEbiography}[{\includegraphics[width=1in,height=1.25in,clip,keepaspectratio]{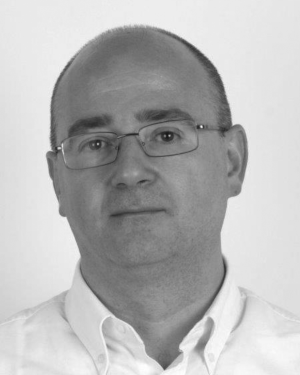}}]{Joan Garcia-Haro}
(M’91) received the M.S and Ph.D degrees in telecommunication engineering from the Universitat Politècnica de Catalunya, Barcelona, Spain, in 1989 and 1995, respectively. He is currently a Professor with the Universidad Politécnica de Cartagena (UPCT). He is author or co-author of more than 70 journal papers mainly in the fields of switching, wireless networking and performance evaluation. Prof. Garcia-Haro served as Editor-in-Chief of the IEEE Global Communications Newsletter, included in the IEEE Communications Magazine, from April 2002 to December 2004. He has been Technical Editor of the same magazine from March 2001 to December 2011. He also received an Honorable Mention for the IEEE Communications Society Best Tutorial paper Award (1995). He has been a visiting scholar at Queen’s University at Kingston, Canada (1991-1992) and at Cornell University, Ithaca, USA (2010- 2011).
\end{IEEEbiography}

\end{document}